\documentclass[12pt]{iopart}
\usepackage{graphicx}
\usepackage{cite}

\begin{document}

\title[Faraday and resonant waves in binary collisionally-inhomogeneous BECs]{Faraday and resonant waves in binary collisionally-inhomogeneous Bose-Einstein condensates}

\author{J. B. Sudharsan$^1$, R. Radha$^1$, Mihaela Carina Raportaru$^{2,3}$, Alexandru I. Nicolin$^{2,3}$ and Antun Bala\v z$^4$}

\address{$^1$Centre for Nonlinear Science (CeNSc), Post-Graduate and Research Department of Physics,
Government College for Women (Autonomous), Kumbakonam 612001, India\\
$^2$Horia Hulubei National Institute of Physics and Nuclear Engineering (IFIN-HH), P.~O.~Box MG-6, 077125 Magurele, Romania\\
$^3$Faculty of Physics, University of Bucharest, Atomistilor 405, Magurele, Romania\\
$^4$Scientific Computing Laboratory, Institute of Physics Belgrade, University of Belgrade, Pregrevica 118, 11080 Belgrade, Serbia}
\ead{mraportaru@nipne.ro}

\begin{abstract}
We study Faraday and resonant waves in two-component quasi-one-dimensional
(cigar-shaped) collisionally inhomogeneous Bose-Einstein condensates
subject to periodic modulation of the radial confinement. We show
by means of extensive numerical simulations that, as the system exhibits
stronger spatially-localised binary collisions (whose scattering length is taken for convenience to be
of Gaussian form), the system becomes effectively a linear one. In other words, as the scattering length approaches a delta-function,
we observe that the two nonlinear configurations typical for binary cigar-shaped condensates,
namely the segregated and the symbiotic one, turn into two overlapping
Gaussian wave functions typical for linear systems, and that the instability onset times
of the Faraday and resonant waves become longer. Moreover, our numerical
simulations show that the spatial period of the excited waves (either resonant
or Faraday ones) decreases as the inhomogeneity becomes stronger.
Our results also demonstrate that the topology of the ground state impacts the dynamics of the ensuing density waves,
and that the instability onset times of Faraday and resonant waves, for a given level of inhomogeneity  in the two-body interactions, depend on whether the initial configuration is segregated or symbiotic.
\end{abstract}


\section{Introduction}

Pattern formation is one of the recurrent research themes which covers
both classical and quantum systems. On the side of classical systems,
it extends over a wide set of topics such as thermal convection in
fluids, Taylor-Couette flows, emergence of patterns in solidification
fronts, chemical reactions and excitable biological media, to name
only the most prominent examples (see \cite{PatternFormationReview}
for a detailed review), while on the side of quantum systems, one of
the most important research directions concerns ultracold quantum
gases.

There is a long list of recent developments on the emergence
and dynamics of nonlinear wave-forms in ultracold quantum
gases both from the theoretical \cite{St1,St2,Da1,Da2} and experimental perspective (see \cite{ReviewRRP, RRRRP} respectively for a comprehensive
treatment of the subject), of which we mention the experiments on
Faraday waves in $^{87}$Rb Bose-Einstein condensates (BECs) \cite{F1} and $^{4}$He cells \cite{F2,F3}, and those on
the collective modes of a $^{7}$Li BEC and its subsequent granulation \cite{F4}.
These experiments paved the way for a series of theoretical investigations
dedicated to the emergence of density waves (e.g., Faraday and resonant)
in condensates with short-range \cite{F5,F6} and dipolar \cite{F7} interactions, in collisionally
inhomogeneous condensates \cite{F8,F9}, Fermi-Bose mixtures, superfluid Fermi gases \cite{F11},
etc., as well as the suppression of Faraday waves and density excitations
in general by means of space- and time-modulated potentials \cite{F12}. The effects of disorder \cite{Dis1,Dis2,Dis3} and interplay with quantum fluctuations \cite{QF1,QF2} still remain to be explored.

To understand the appeal that pattern-forming modulational instabilities and the ensuing density waves have exerted in the ultracold gases community, one has to contrast the almost unprecedented level of experimental control seen in ultracold gases with that commonly seen in classical fluids. In a way, ultracold gases in general and BECs in particular, have become the ideal testbed for nonlinear waves due to the control over the geometry of the experimental setup (through various magnetic and/or optical traps), the tunability of the effective nonlinearity of the system (using the magnetic and optical Feshbach resonances detailed below) and the accuracy of the Gross-Pitaevskii equation (GPE) which describes
the dynamics of the condensate at zero temperature.

Thanks to the experimental development of techniques for magnetic and
optical Feshbach resonances, it was possible to probe the so-called
collisionally inhomogeneous regime (a term coined in \cite{CIC}),
which is characterised by spatial variations in the strength of the
two-body interactions. Magnetic Feshbach resonances have a longer
history than their optical siblings, with significant experimental
results on the formation of ultracold molecules, the BEC-BCS crossover,
and the production of Efimov trimer states. Despite these results,
this method is not particularly useful to reach the collisionally
inhomogeneous regime because the length scale for application of the
Feshbach field is usually larger than the characteristic size of the
condensate. Optical Feshbach resonances have been shown, however,
to generate spatial variations of the scattering length on the scale
of hundred nanometers (see, for instance, \cite{OpticalFeshbach}) and
are, therefore, the preferred method of choice. Among the 
numerous experimental protocols used to reach the collisionally inhomogeneous 
regime, we refer to that described in \cite{OpticalFeshbach2} where it was shown that under 
specific experimental conditions, a Gaussian optical field can impart a 
similar spatial profile to the two-body interaction.

In this paper we focus on pattern-forming modulational instabilities in binary cigar-shaped collisionally-inhomogeneous Bose-Einstein condensates and show by extensive numerical calculations that, as the binary collisions get localised at the centre of the underlying magnetic trap, the system reaches an effectively linear regime. The two non-miscible configurations typical for binary cigar-shaped condensates (namely the segregated and the symbiotic) now turn into a miscible configuration in which the two components overlap with each other and the individual wave functions are close to two Gaussians. Moreover, the excitation of density waves of Faraday and resonant type by means of periodic modulation of the radial component strength of the magnetic trap is substantially slowed down, as one can easily see from the instability onset times. Due to the complex structure of the wave function, we observe that the usual variational approach provides very limited analytical insight into the dynamics of the system and therefore report only numerical results obtained from the Gross-Pitaevskii equation. The rest of the manuscript is structured as follows: in section \ref{numerics} we present the numerical treatment of the Gross-Pitaevskii equation, in section \ref{statics} we show the stationary configurations of the system and the transition to a miscible configuration, while in section \ref{dynamics} we present in detail the dynamics of the condensate. Finally, in section \ref{conclusions} we gather our concluding remarks.

\section{Mean-field theory and numerical approach}
\label{numerics}

Many of the theoretical investigations into the properties of BECs
mentioned in the previous section rely on an accurate numerical
treatment of the mean-field GPE. For the ground state of a two-component BEC system, it reads
\begin{equation}
\mu_j \psi_j = -\frac{1}{2}\Delta \psi_j + V({\bf r}, t) \psi_j + N_j G_j({\bf r}) \vert \psi_j \vert^2 \psi_j +
N_{3-j}G_{12}({\bf r}) \vert \psi_{3-j} \vert^2 \psi_j\, ,
\label{GPET1}
\end{equation}
while the dynamics is determined by
\begin{equation}
i \frac{\partial \psi_j}{\partial t}  =  -\frac{1}{2}\Delta \psi_j + V({\bf r}, t) \psi_j + N_j G_j({\bf r}) \vert \psi_j \vert^2 \psi_j +
N_{3-j}G_{12}({\bf r}) \vert \psi_{3-j} \vert^2 \psi_j\, 
\label{GPET2}
\end{equation}
where $j\in\{ 1, 2\}$, and $N_j$ is the fixed number of atoms in the component $j$. The number of atoms is fixed in equations by introducing the Lagrange multipliers $\mu_j$, which represent the nonlinear analogues of the eigenenergies.
For simplicity, we use natural units $\hbar = m = 1$ in all equations shown throughout the paper, and the component wavefunction normalisation is taken to be
\begin{equation}
\int d{\bf r} \vert \psi_j({\bf r}, t) \vert^2 = 1\, .
\label{NORM}
\end{equation}
The strength of the nonlinearities $G_j({\bf r})$ and $G_{12}({\bf r})$ are proportional to the corresponding intra- and inter-component $s-$wave scattering lengths, which can be engineered to be spatially inhomogeneous using optical Feshbach resonances (as was demonstrated in \cite{OpticalFeshbach2}).

In this paper, we consider two hyperfine states of $^{87}$Rb (referred here as states A and B) in an external harmonic trapping potential of the form
\begin{equation}
V({\bf r}, t) = \frac{1}{2} \Omega_\rho^2(t) \rho^2 + \frac{1}{2} \Omega_z^2 z^2\, ,
\label{POTFORM}
\end{equation}
where $\rho^2 = x^2 + y^2$, such that the system and component wave functions are cylindrically symmetric, i.e., $\psi_j({\bf r}, t) \equiv \psi_j(\rho, z, t).$ We also assume that the system is strongly confined in radial direction, i.e., $\Omega_\rho(t) \gg \Omega_z$ and that the scattering length is spatially modulated only in the radial direction, such that the nonlinear interactions have the form
\begin{equation}
G_j({\bf r})=G_j(\rho) = 4 \pi a_j(0)\, e^{-\rho^2 / 2 b^2} = g_j\, e^{-\rho^2 / 2 b^2},
\label{SCLENa}
\end{equation}
\begin{equation}
G_{12}({\bf r})=G_{12}(\rho) = 4 \pi a_{12}(0)\, e^{-\rho^2 / 2 b^2} = g_{12}\, e^{-\rho^2 / 2 b^2},
\label{SCLENb}
\end{equation}
where $a_j(0)$ and $a_{12}(0)$ represent the constant $s$-wave scattering lengths for intra- and inter-component collisions respectively along the $z$-axis, and $b$ is the length scale of the spatial modulation of the scattering length in the radial direction. Here, we use the experimentally observed values for the scattering lengths
from \cite{verhaar,middlekamp,hamner}:
\begin{equation}
a_1(0) = 100.4\, a_0\, ,\quad a_2(0) = 98.98\, a_0\, ,\quad a_{12}(0) = a_1(0)
\label{SCVAL}
\end{equation}
where $a_0$ is the Bohr radius.

The numerical treatment of the GPE gradually developed into
a research direction in its own right and now fast and accurate
numerical algorithms exist for calculation of the ground and excited states of BECs
using imaginary-time propagation \cite{ImaginaryTimeAlgorithms}, as well as
explicit finite-difference scheme \cite{ExplicitFiniteDifferenceScheme},
time-splitting spectral methods \cite{TimeSplittingSpectralMethods},
methods based on expansion of the condensate wave function in terms
of the solutions of the harmonic oscillator which characterises
the magnetic trap \cite{BasisSet}, symplectic shooting method \cite{Symplectic},
etc. A popular package of codes (available in Fortran and C \cite{GP1,GP2,GP3,GP4,GP5,GP6}
programming languages, including parallelised versions in MPI and CUDA) has proven to be particularly useful, as
it provides the stationary states and the nonlinear dynamics of one-,
two- and three-dimensional BECs. The C codes, in particular, are OpenMP-parallelised
such that the execution time decreases substantially (typically by an order
of magnitude) compared to serial ones, even on a modern desktop computer, when all available CPU cores are used by the programme.

In our {\it in silico} experiments, we use the adapted codes from \cite{GP2} for a system with $N_1=2.5\times 10^{5}$ atoms in
the state A and $N_2 = 1.25\times 10^5$ atoms in the state B, loaded into a
quasi-one-dimensional magnetic trap with frequencies $\Omega_{\rho0}=160\times 2\pi$ Hz
and $\Omega_{z}=7\times 2\pi$ Hz. The ground state of the system is
computed numerically using the method of imaginary-time
propagation considering the scattering lengths given in
equation (\ref{SCVAL}). This method relies on the change of
variable $t=-i\tau$, which transforms the GPE into a nonlinear
diffusion equation. Using the standard renormalisation of wavefunctions to unity after each
time step, which is necessary since the imaginary-time propagation does not conserve unitarity, we converge to a ground state of the system
whose energy is a local minimum.

At first, we determine the ground state of a single condensate component with the constant $s$-wave scattering length
$a = 100.0\, a_0$, and calculate its radial width. For this quantity, we obtain the value $b_0 = 1.86$  $\mu m$, which then serves as
a reference length scale for expressing the value of the inhomogeneity
parameter $b$ in equations (\ref{SCLENa}) and (\ref{SCLENb}).
Afterwards, we turn to the full two-component system with the interaction parameters specified in equation (\ref{SCVAL}),
calculate its ground state and study the dynamics for various values of the inhomogeneity parameter $b$.
For each configuration under consideration, we numerically observe the dynamics of the
condensate subject to the parametric modulation of the radial trap frequency
$\Omega_{\rho}(t)=\Omega_{\rho0}(1+\epsilon \sin\omega t)$, where
$\epsilon$ and $\omega$ represent the modulation amplitude and
the corresponding frequency.

\section{Stationary configurations}
\label{statics}

Two-component BEC systems with short-range contact interactions can be either miscible or non-miscible, depending on the relationship between their intra- and inter-component scattering lengths. In general, if the condition $g_1 g_2<g_{12}^2$ is satisfied, the system turns to be non-miscible, with the two components clearly separated \cite{miscibility,ref50A,ref38A,Ivana}. Otherwise, the system is miscible, and the components have a significant overlap. This is possible only when the inter-component interaction is smaller than the intra-component ones, which means that mixing of the components is energetically favoured. In our case, the above condition is satisfied, and the system is expected to be non-miscible.

When such a system is loaded into a quasi-one-dimensional trap, it can exhibit one of two possible types of stationary non-miscible
configurations \cite{FW2012}: a segregated one, in which the two components face each other, and a symbiotic one, in which one component
effectively traps the other, forming a structure akin to a
bright-dark soliton molecule. Depending on the interaction strengths and their relationship, one of these configurations is a ground state, and the other is an excited state. For the system we are considering, the symbiotic pair represents a ground state, and the segregated state is an excited state. Using various initial conditions and the methods described above, we are able to numerically compute both the above mentioned non-miscible configurations.

Moreover, as we will see in this section, our numerical results show that for a
two-body scattering length modulated spatially, the aforementioned
non-miscible configurations can change their character and become miscible. In particular, as
the scattering length gets localised around the centre of the
underlying magnetic trap, the two components of the condensate
start to have significant overlap and eventually reach a miscible state (in which
the wavefunctions of both components are very close to a
Gaussian) when the spatial profile of the scattering
length reaches a delta-function-like profile. To understand this
transition to miscibility, we should recall that as the scattering
length gets more and more localised, i.e., the parameter $b$ decreases, the effective nonlinearity of the system
decreases as well. This is transparent from the expressions for the effective nonlinearities:
\begin{equation}
\tilde{G}_j = \int_0^{\infty} d \rho\, 2 \pi \rho G_j(\rho) = 2 \pi g_j b^2\, ,
\label{EFFGa}
\end{equation}
\begin{equation}
\tilde{G}_{12} = \int_0^{\infty} d \rho\, 2 \pi \rho G_{12}(\rho) = 2 \pi g_{12} b^2\, .
\label{EFFGb}
\end{equation}
Please note that the effective nonlinear interaction depends quadratically on $b$. This means that for smaller values of $b$, the nonlinear effect fades out and the condensate reaches an effective linear regime.

\subsection{Symbiotic pair state - ground state}

\begin{figure}[!ht]
\begin{center}
\includegraphics[width=70 mm]{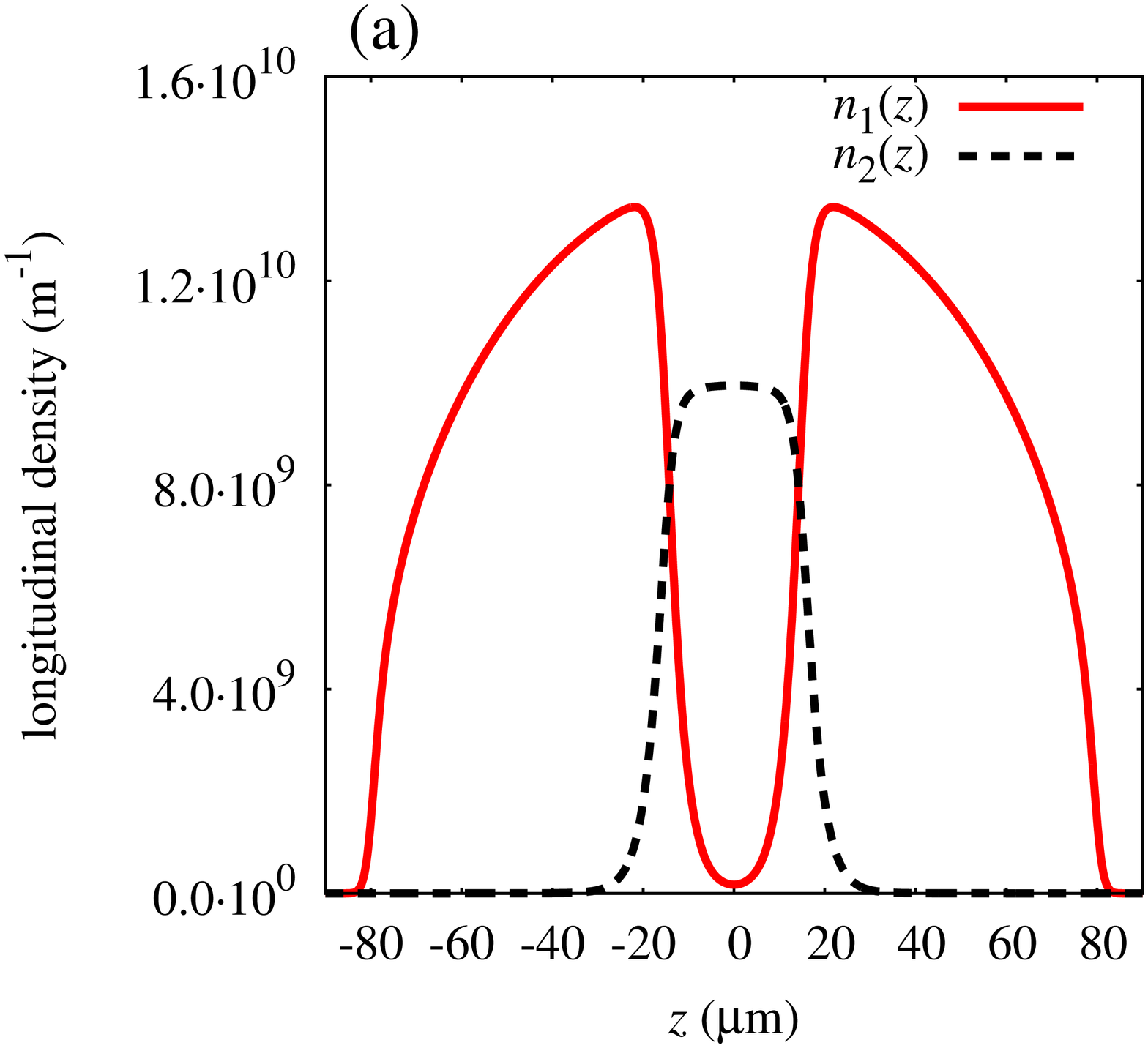}
\includegraphics[width=70 mm]{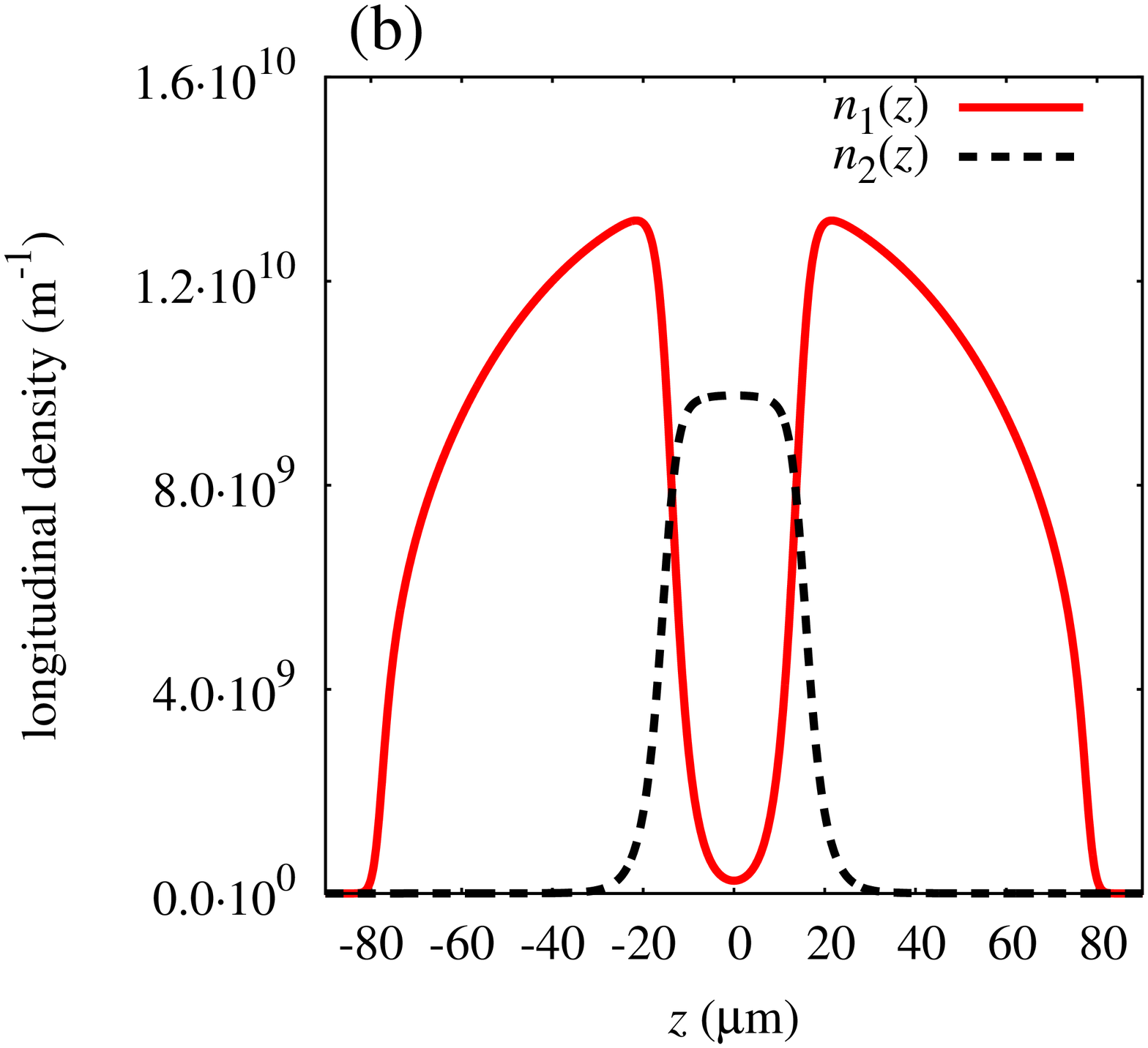}\\
\includegraphics[width=70 mm]{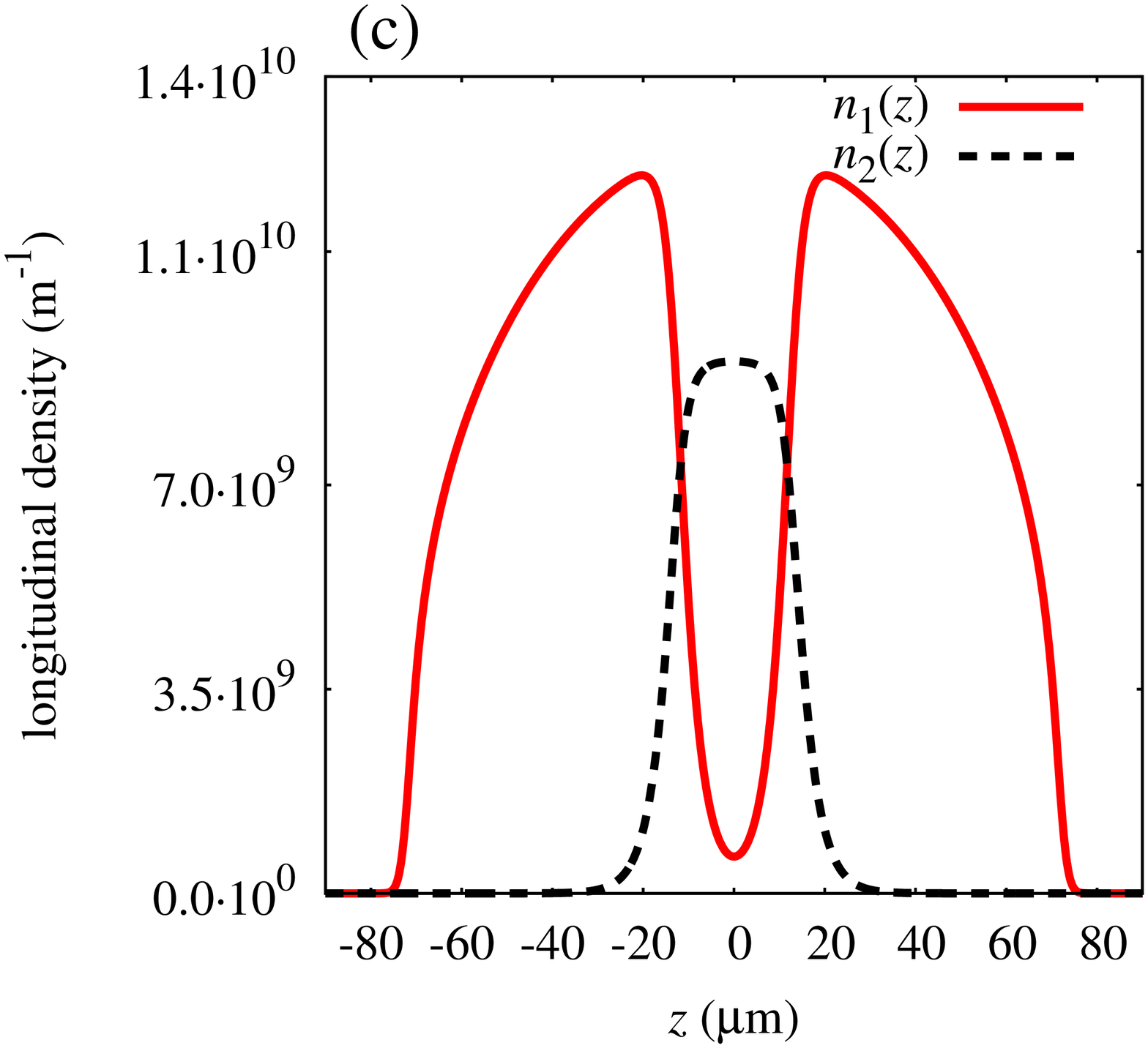}
\includegraphics[width=70 mm]{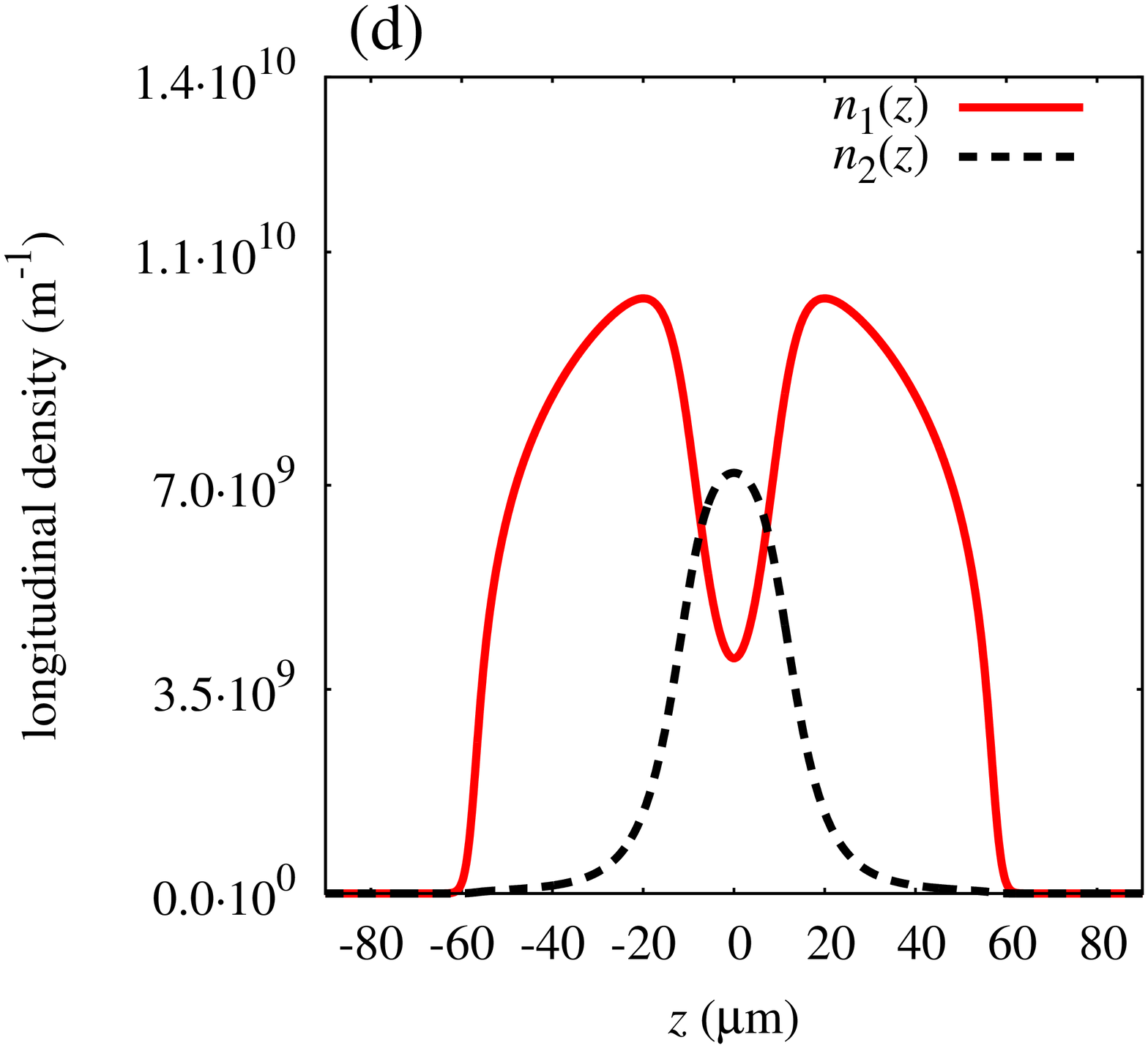}\\
\includegraphics[width=70 mm]{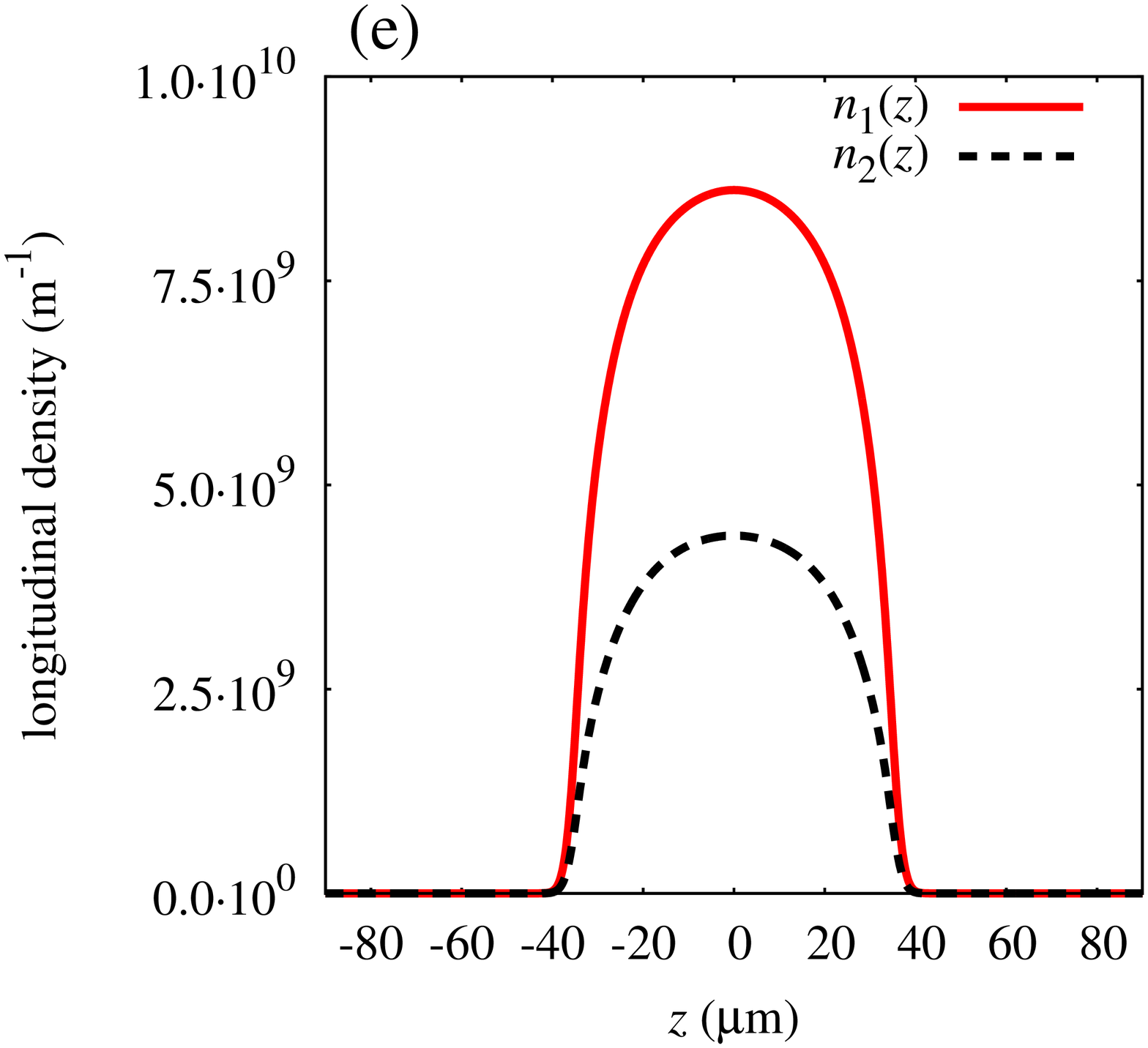}
\caption{Ground states - the symbiotic pair stationary states - of a binary condensate in the collisionally inhomogeneous regime. The longitudinal density profile $n_1(z)$ and $n_2(z)$ are depicted for various values of the inhomogeneity control parameter $b$: (a) $b = 4b_0$, (b) $b=2b_0$, (c) $b=b_0$, (d) $b=b_0/2$, and (e) $b=b_0/4$. As the inhomogeneity parameter decreases, the overlap of the two components increases, and the system gradually transforms into a miscible one.}
\label{Sym-Stationary}
\end{center}
\end{figure}

The symbiotic pair stationary state solution consists of one component well localised in the
centre of the harmonic trap, while the other component surrounds it. Figure \ref{Sym-Stationary} shows the
radially integrated density profiles of the two components,
\begin{equation}
n_j(z) = \int_0^{\infty} d\rho\, 2 \pi \rho | \psi_j(\rho, z) |^2\, ,
\label{RADIAL-INTE}
\end{equation}
obtained by imaginary-time propagation, starting from two identical Gaussian profiles (see equation (\ref{IDG})) for various strengths of inhomogeneity control parameter $b$,
 \begin{equation}
 \psi_1(\rho, z, t=0) = \psi_2(\rho, z, t=0) = \frac{1}{\pi^{3/2}}\, e^{ - \frac{1}{2} (\rho^2 + z^2)}\, .
 \label{IDG}
 \end{equation}
One can clearly observe that the immiscible configuration of the two components disappears as  we decrease the strength of the inhomogeneity parameter $b$, and that the two components become miscible for small enough $b$, as expected from the above consideration of effective interactions.

\subsection{Segregated state}
\label{sec:seg-stationary}

\begin{figure}[!ht]
\begin{center}
\includegraphics[width=70 mm]{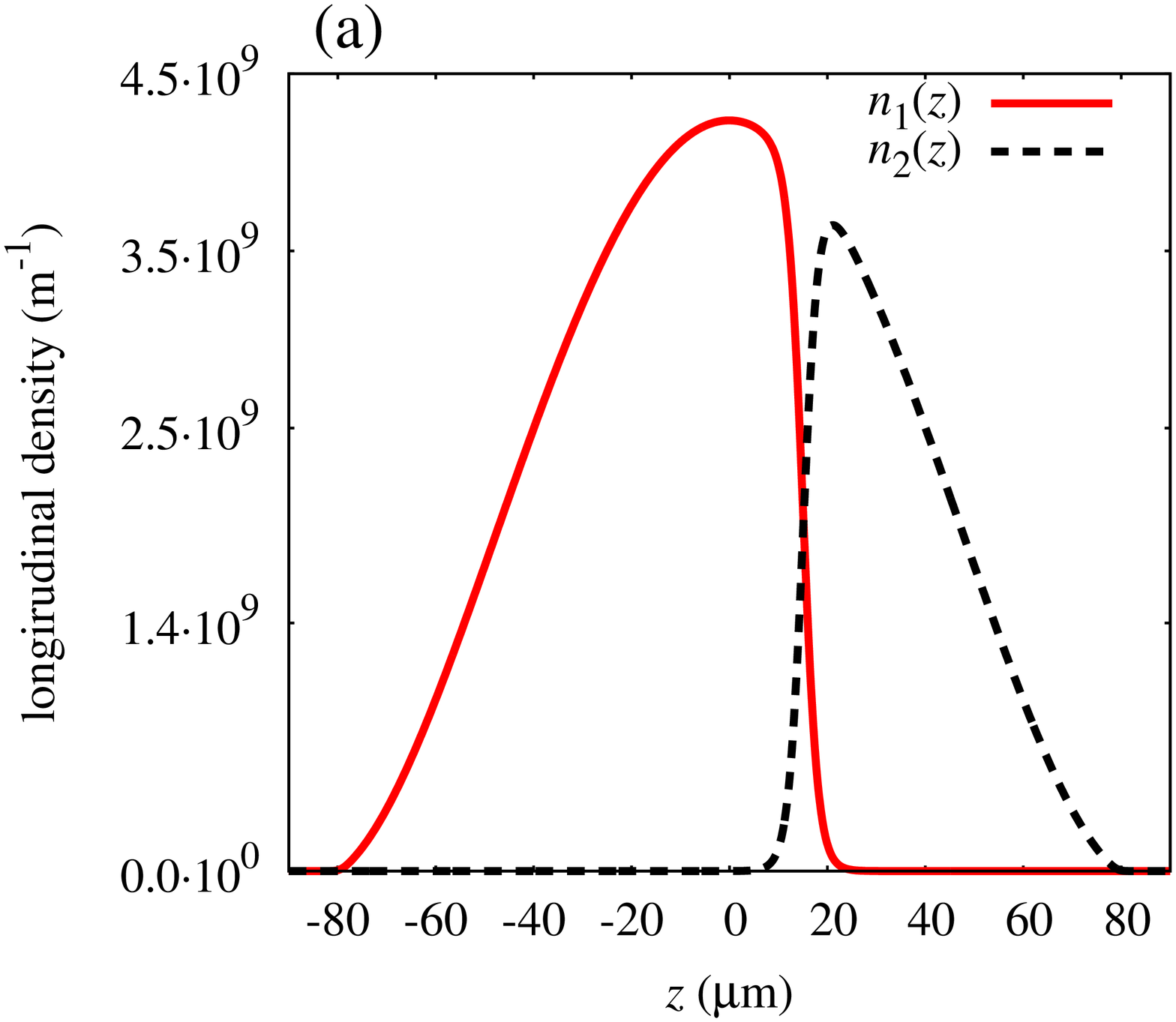}
\includegraphics[width=70 mm]{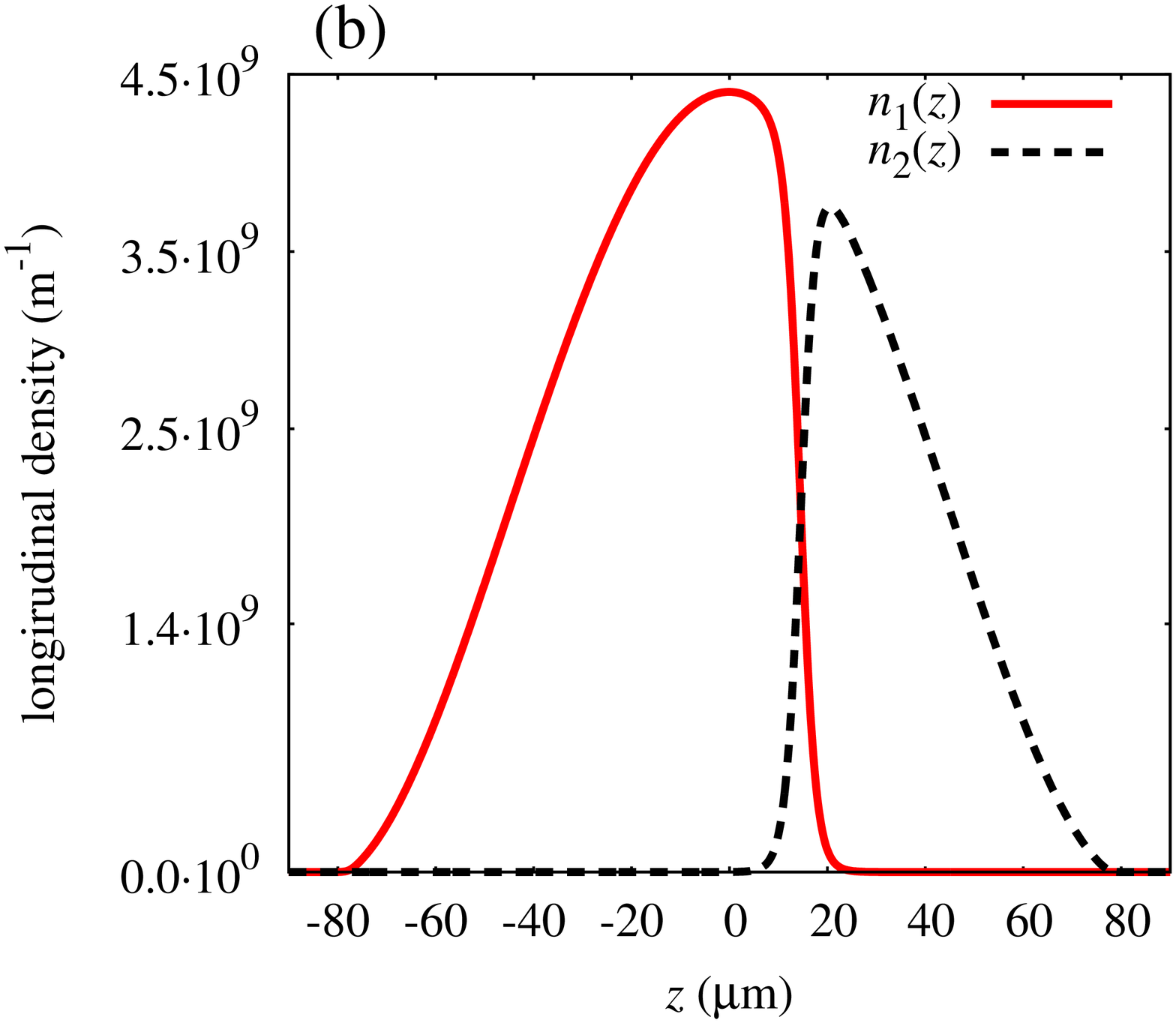}\\
\includegraphics[width=70 mm]{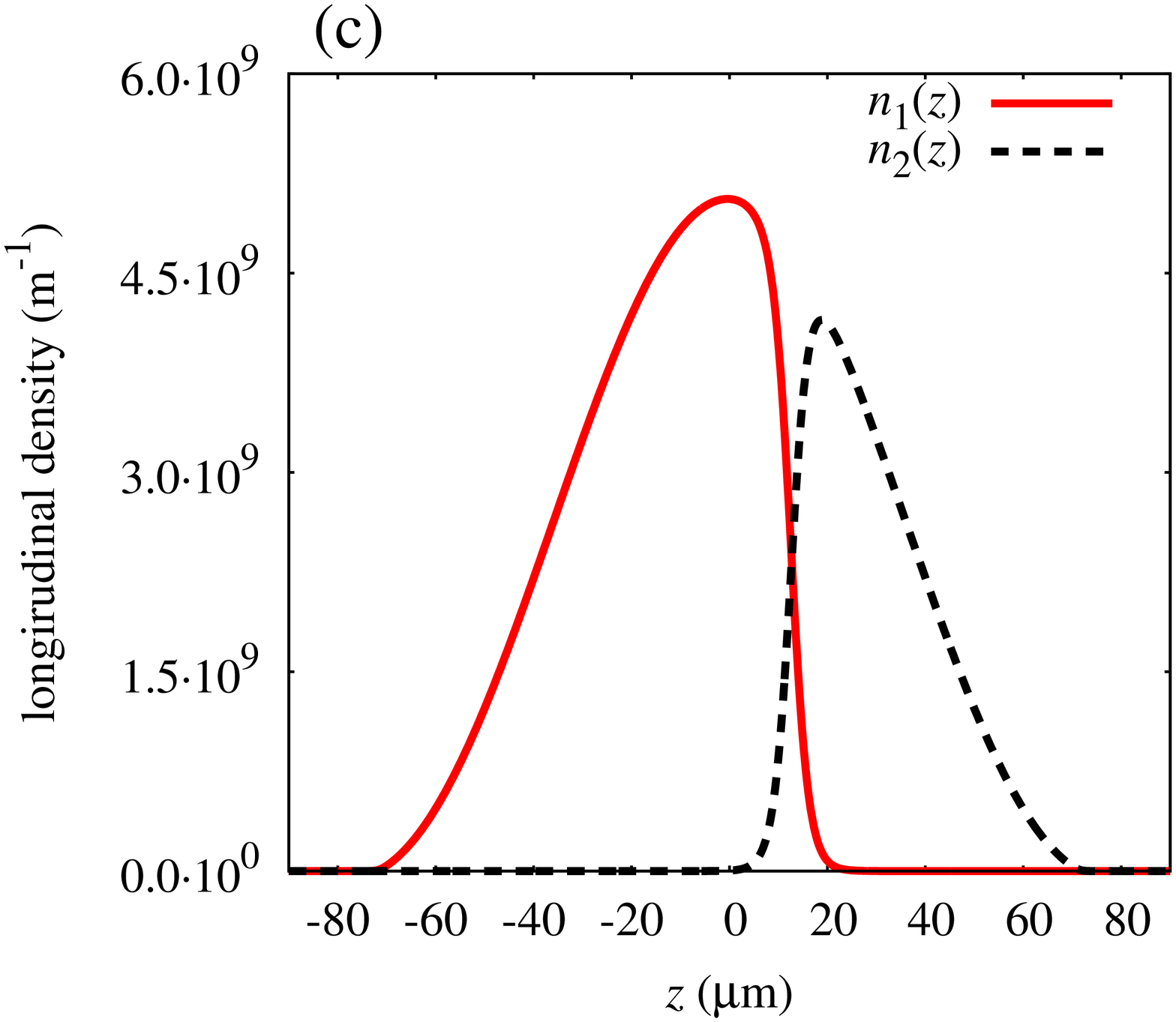}
\includegraphics[width=70 mm]{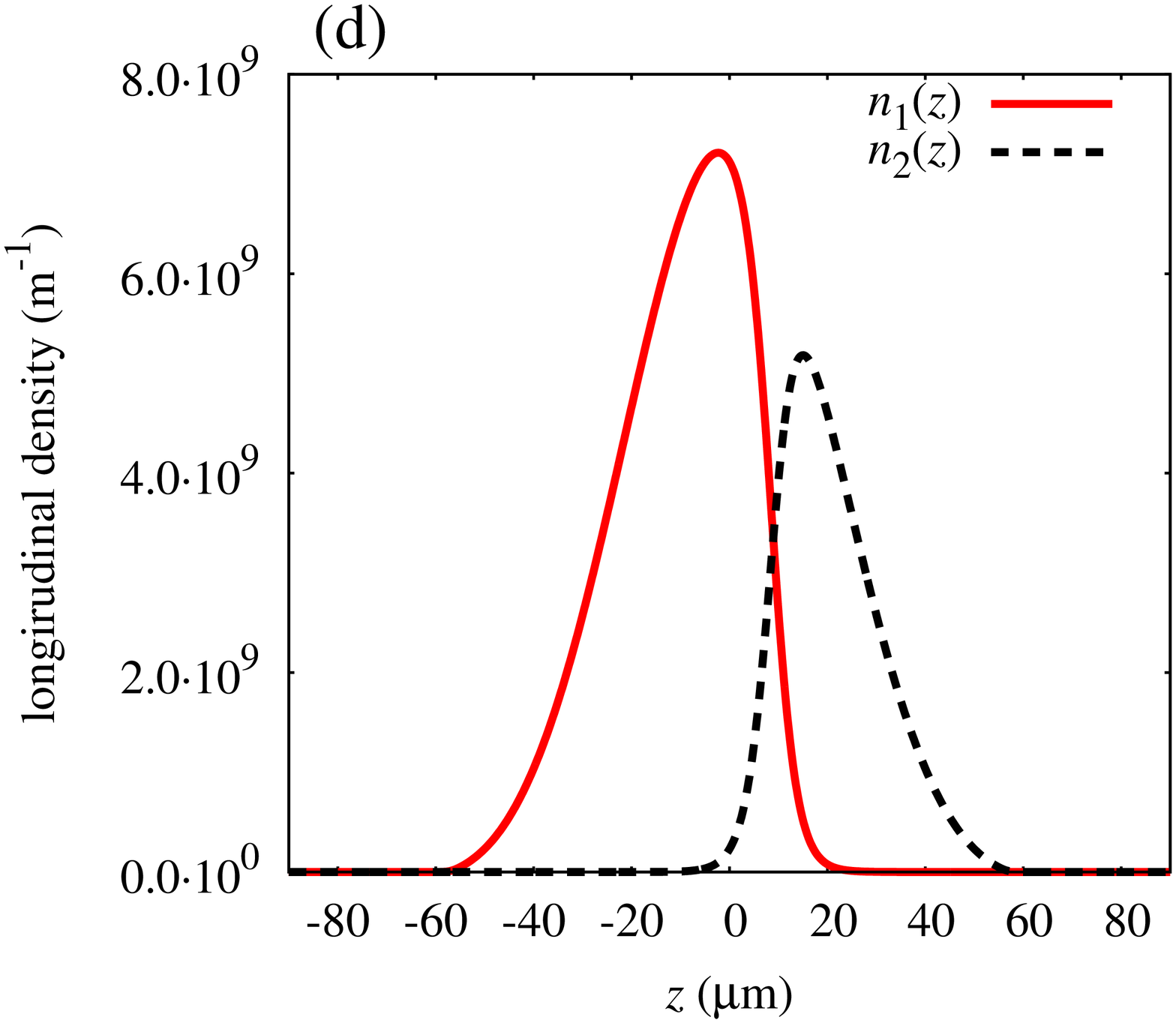}\\
\includegraphics[width=70 mm]{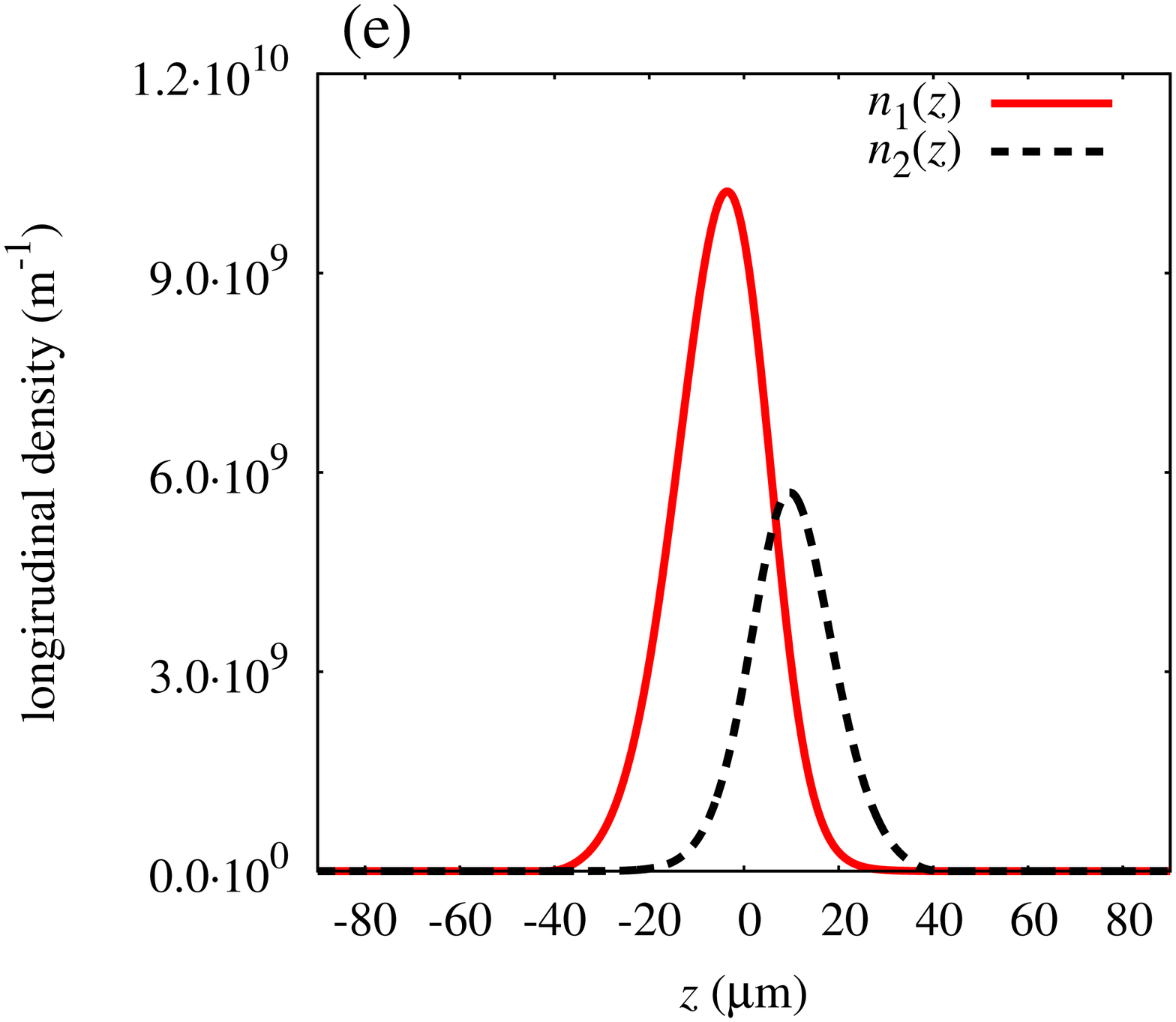}
\caption{The segregated stationary states of a binary condensate in the collisionally inhomogeneous regime. The longitudinal density profile $n_1(z)$ and $n_2(z)$ are depicted for various values of the inhomogeneity control parameter $b$: (a) $b = 4b_0$, (b) $b=2b_0$, (c) $b=b_0$, (d) $b=b_0/2$, and (e) $b=b_0/4$. As the inhomogeneity parameter decreases, the overlap of the two components increases, and the system gradually transforms into a miscible one.}
\label{Seg-Stationary}
\end{center}
\end{figure}

Figure \ref{Seg-Stationary} shows the radially integrated density profiles for the segregated state obtained by imaginary-time propagation starting from the two well-separated Gaussians,
\begin{equation}
\psi_1(\rho, z, t=0) = \frac{1}{\pi^{3/2}}\, e^{- \frac{1}{2} \left[ (\rho - \rho_0)^2 + (z-z_0)^2\right]}\, ,
\label{NIDGa}
\end{equation}
\begin{equation}
\psi_2(\rho, z, t=0) = \frac{1}{\pi^{3/2}}\, e^{- \frac{1}{2} \left[ (\rho + \rho_0)^2 + (z+z_0)^2\right]}\, .
\label{NIDGb}
\end{equation}

Figures \ref{Seg-Stationary}(a)-(e) show the dependence of the segregated ground state on the value of the inhomogeneity parameter $b$. Again, one notices the disappearance of the immiscible nature of the condensate as we decrease the value of $b$, i.e., as we make the localisation of the interaction around the $z$-axis stronger.

Comparing figures \ref{Sym-Stationary} and \ref{Seg-Stationary}, we observe that in the symbiotic state, the maximal densities of both components are larger for the weak inhomogeneity (large $b$) than for the strong inhomogeneity (small $b$). The main reason for this is that in the case of weak inhomogeneity the two components are well separated, and thus individually have smaller amount of effective available physical space to occupy. As the inhomogeneity increases, the components become more miscible, the amount of effective available physical space per component increases, and the maximal densities become smaller. In the case of segregated stationary state, we observe a reversal of the above phenomenon that can be attributed to a different shape of the two components and the squeezing due to the increasing inhomogeneity (smaller $b$), which in this case makes the effective available physical space smaller as the inhomogeneity increases.

Although in general the condition $g_1 g_2 > g_{12}^2$ is required for the two components to be miscible \cite{miscibility,ref50A,ref38A,Ivana}, and this is not satisfied in our case, we see that spatially inhomogeneous interactions can be suitably exploited, thus making the components miscible in the case of strong inhomogeneity. Therefore, engineering of spatially inhomogeneous interactions offers a prospect to further control behaviour of binary BEC systems and the level of miscibility of the components.

\section{Dynamical results}
\label{dynamics}

\begin{figure}[!ht]
\begin{center}
\includegraphics[width=70 mm]{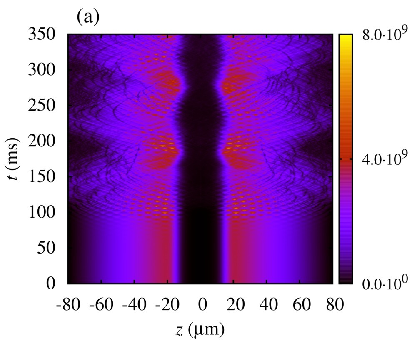}
\includegraphics[width=70 mm]{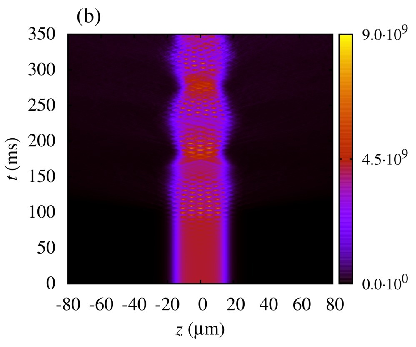}\\
\includegraphics[width=70 mm]{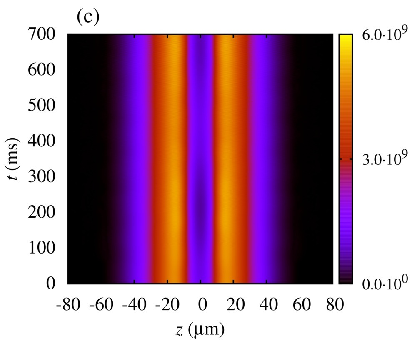}
\includegraphics[width=70 mm]{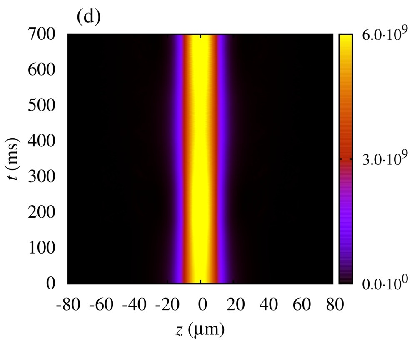}\\
\includegraphics[width=70 mm]{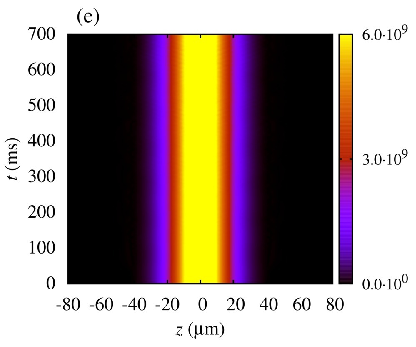}
\includegraphics[width=70 mm]{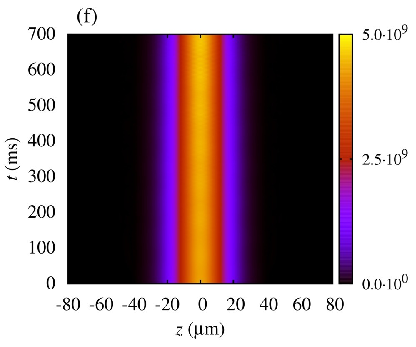}
\caption{Real-time evolution of radially integrated density profiles for weakly and strongly inhomogeneous interactions in the case of a symbiotic configuration
for $\omega = \omega_\mathrm{res}$. The panels on the left (right) correspond to the component A (B) for the inhomogeneity parameter
values: (a) and (b) $b = 4 b_0$, (c) and (d) $b = b_0/2$, (e) and (f) $b = b_0/4$. Note the softening of nonlinear excitations  as $b$ decreases and the system reaches an effectively linear regime}
\label{160-ws-psi12}
\end{center}
\end{figure}

\begin{figure}[!ht]
\begin{center}
\includegraphics[width=70 mm]{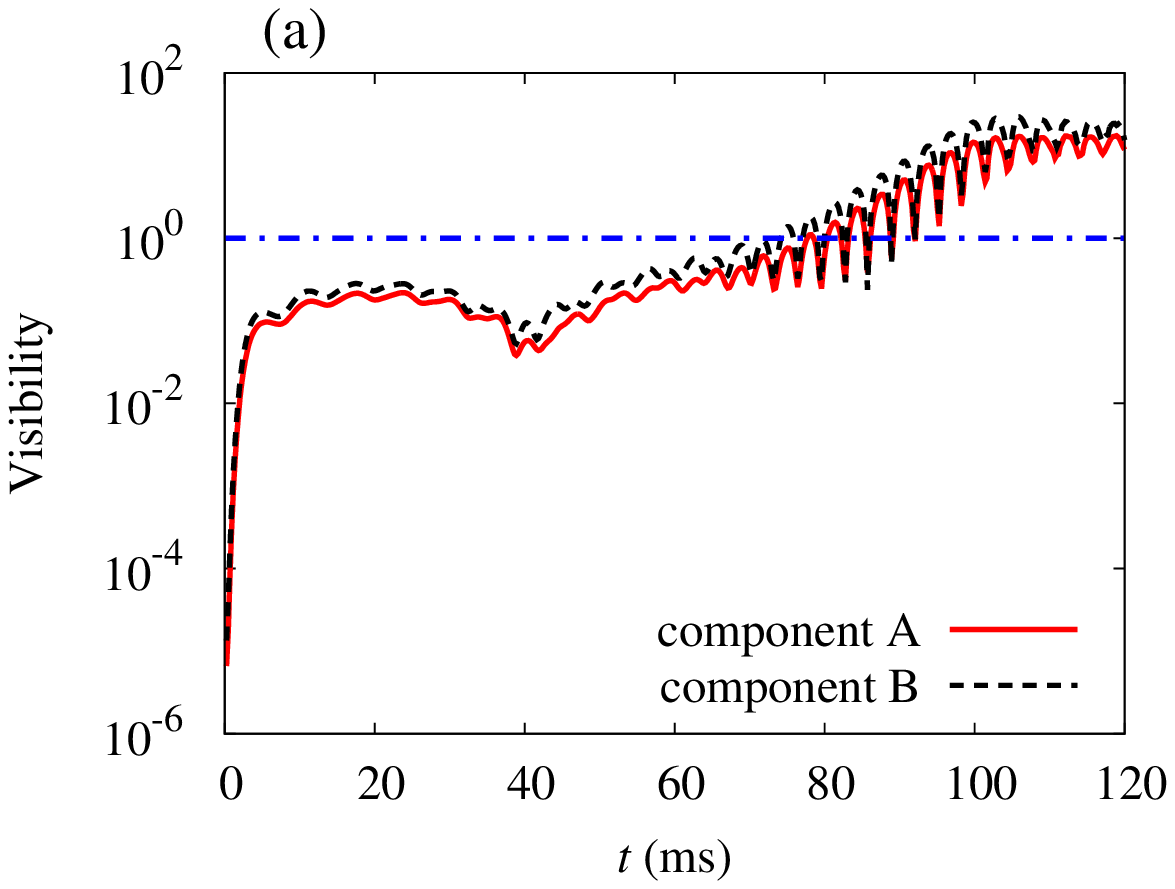}
\includegraphics[width=70 mm]{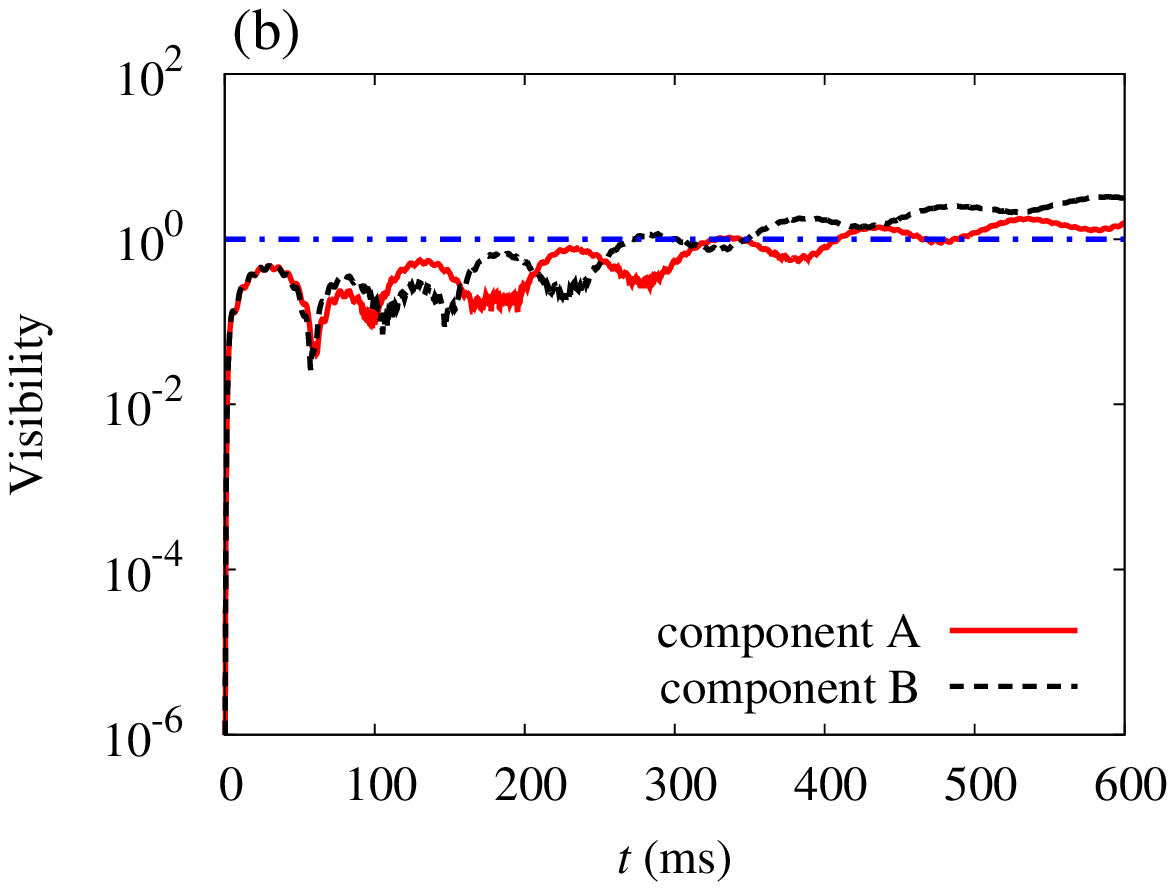}
\caption{Time dependence of the visibility during real-time evolution for weakly and strongly inhomogeneous interactions in the case of a symbiotic configuration
for $\omega = \omega_\mathrm{res}$ for the inhomogeneity parameter values: (a) $b = 4 b_0$, (b) $b = b_0/4$. The horizontal dashed-dotted line corresponds to the
visibility equal to one and denotes the onset of resonant waves.}
\label{160-ws-vis}
\end{center}
\end{figure}

\begin{figure}[!ht]
\begin{center}
\includegraphics[width=70 mm]{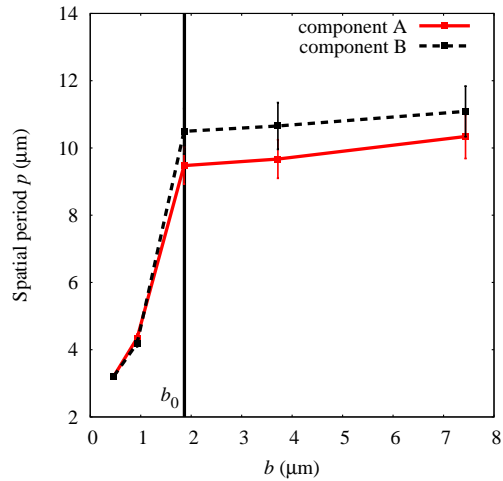}
\caption{Spatial period of resonant waves as a function of inhomogeneity $b$ in the case of a symbiotic configuration for $\omega = \omega_\mathrm{res}$,
obtained using FFT analysis.}
\label{160-period}
\end{center}
\end{figure}

In this section, we study excitations of the system induced by a harmonic modulation of the radial part of the trapping potential. Such a modulation generates density waves, which can have a form of Faraday or resonant waves \cite{F5,F6,F8}. Faraday waves are characterised by a frequency which is equal to half that of the driving frequency, while the resonant waves have the frequency equal to the driving one. Additionally, the amplitude of a resonant wave increases exponentially, fuelled by an efficient, resonant energy transfer. Here we study how spatially inhomogeneous interactions affect properties of Faraday and density waves in a two-component BEC systems.

We present results for the real-time dynamics of a collisionally inhomogeneous binary condensate with
$N_1 = 2.5 \times 10^5$ atoms of $^{87}$Rb in the hyperfine state A and $N_2 = 1.25 \times
10^5$ atoms of $^{87}$Rb in the hyperfine state B, as for the calculation of stationary configurations.
For both symbiotic and segregated state configurations, we have the harmonic magnetic trap of the form (\ref{POTFORM}), with the parameters
$\Omega_{\rho}(t)=\Omega_{\rho0}(1+\epsilon \sin\omega t)$, 
$\{ \Omega_{\rho 0}, \Omega_{z} \} = \{ 160 \times 2 \pi ,\,  7 \times 2 \pi \}$~Hz \cite{F1}, and two typical modulation (driving) frequencies,
$\omega=\omega_\mathrm{res} = \Omega_{\rho 0}$ and $\omega=\omega_\mathrm{F} = 250 \times 2 \pi$~Hz.
The first driving frequency $\omega_\mathrm{res}$ is equal to the radial frequency of the underlying trap and gives rise to density waves of the same frequency as that of the drive, which will turn out to be resonant waves. The second chosen driving frequency $\omega_\mathrm{F}$ was selected to be strongly off-resonance with both the radial frequency of the trap ($\Omega_{\rho 0}$) and its first harmonic ($2\Omega_{\rho 0}$). This off-resonant drive gives rise to waves of a frequency equal to half that of the drive, commonly known as Faraday waves \cite{St1,St2,RRRRP,F5}.
The modulation amplitude is always set to $\epsilon = 0.1$.

\subsection{Symbiotic pair state}

Starting from the symbiotic ground state solution, we have generated both resonant and Faraday waves for various values of the inhomogeneity parameter $b$.
Figure \ref{160-ws-psi12} shows real-time evolution of radially integrated longitudinal density profiles for the case when resonant waves are obtained, i.e., when the resonant modulation frequency $\omega = \omega_\mathrm{res}$ was used, for both weak (large values of $b$) and strong (small values of $b$) inhomogeneous collisions, respectively.

\begin{figure}[!ht]
\begin{center}
\includegraphics[width=70 mm]{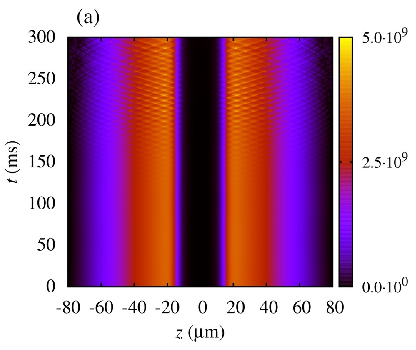}
\includegraphics[width=70 mm]{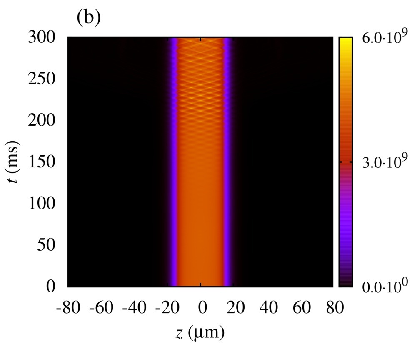}\\
\includegraphics[width=70 mm]{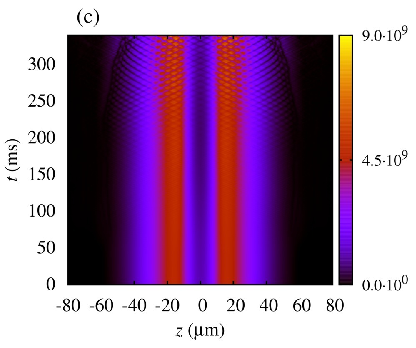}
\includegraphics[width=70 mm]{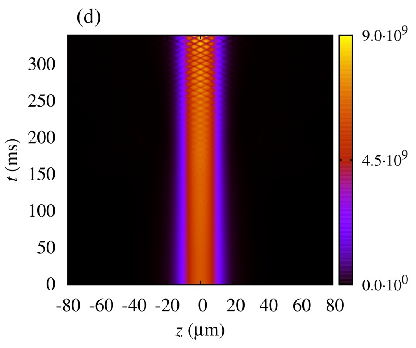}\\
\includegraphics[width=70 mm]{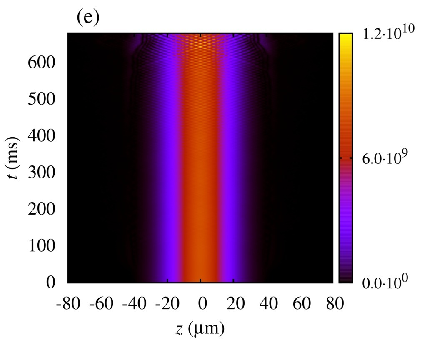}
\includegraphics[width=70 mm]{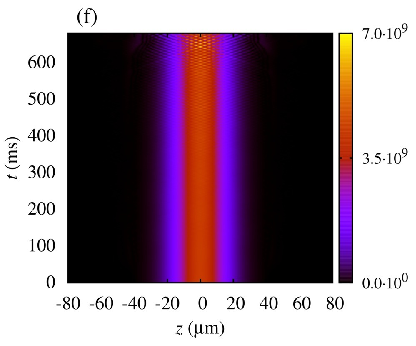}
\caption{Real-time evolution of radially integrated density profiles for weakly and strongly inhomogeneous interactions in the case of a symbiotic configuration
for $\omega = \omega_\mathrm{F}$. The panels on the left (right) correspond to the component A (B) for the inhomogeneity parameter
values: (a) and (b) $b = 4 b_0$, (c) and (d) $b = b_0/2$, (e) and (f) $b = b_0/4$. Note the softening of nonlinear excitations  as $b$ decreases and the system reaches an effectively linear regime.}
\label{250-ws-psi12}
\end{center}
\end{figure}

\begin{figure}[!ht]
\begin{center}
\includegraphics[width=70 mm]{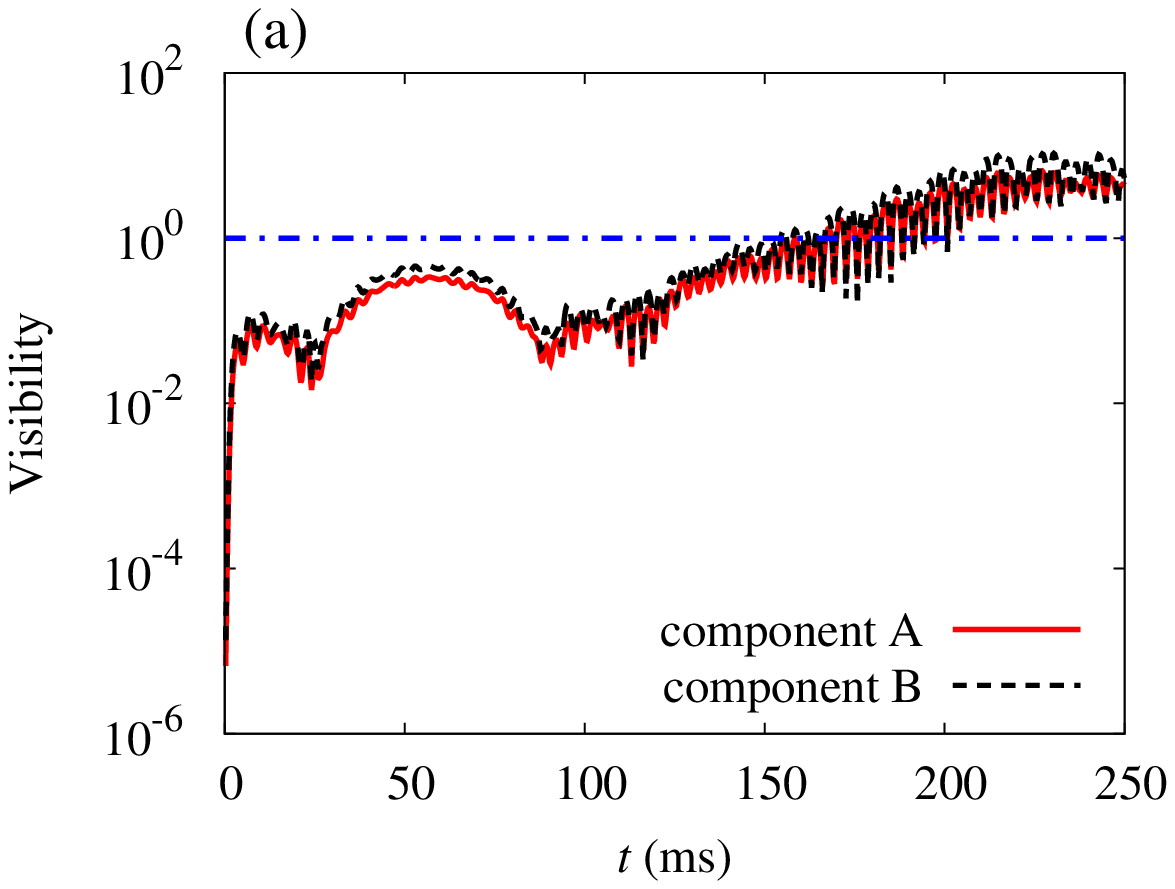}
\includegraphics[width=70 mm]{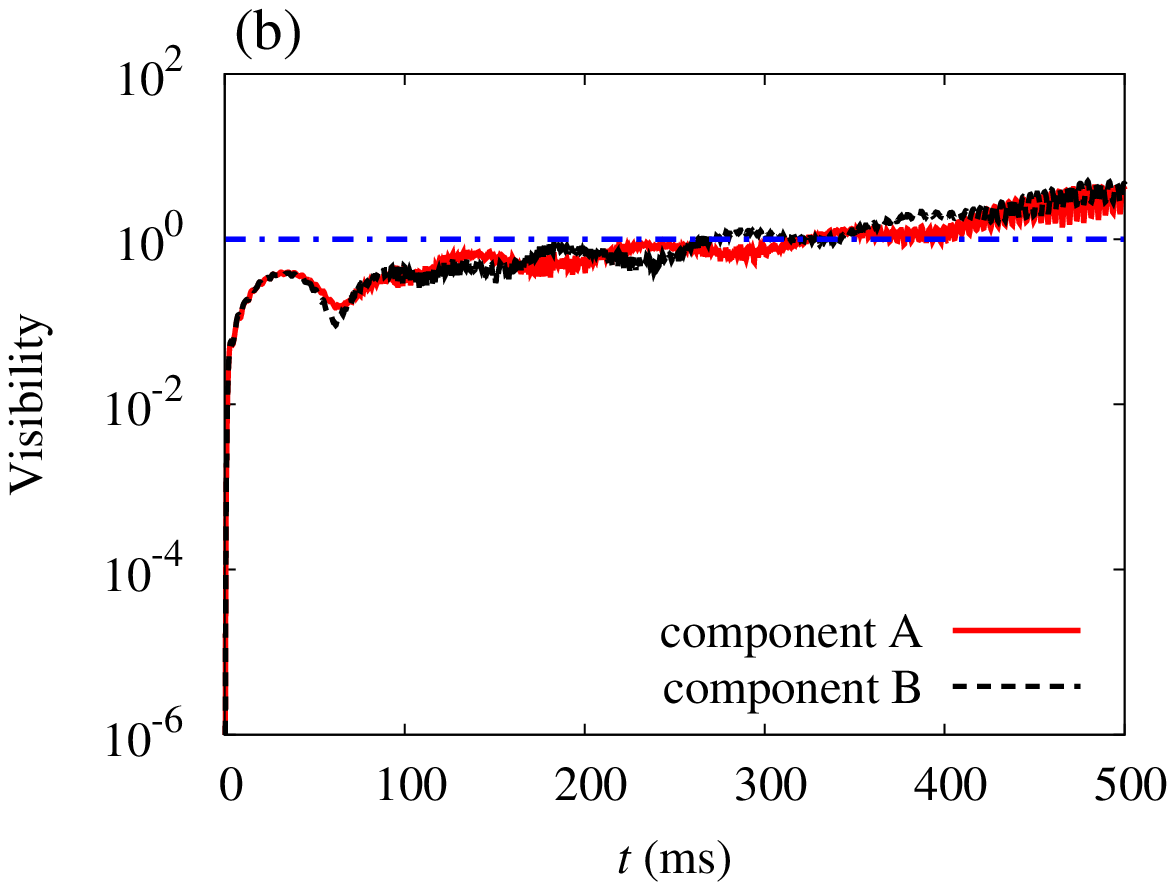}
\caption{Time dependence of the visibility during real-time evolution for weakly and strongly inhomogeneous interactions in the case of a symbiotic configuration
for $\omega = \omega_\mathrm{F}$ for the inhomogeneity parameter values: (a) $b = 4 b_0$, (b) $b = b_0/4$. The horizontal dashed-dotted line corresponds to the
visibility equal to one and denotes the onset of Faraday waves.}
\label{250-ws-vis}
\end{center}
\end{figure}

\begin{figure}[!ht]
\begin{center}
\includegraphics[width=70 mm]{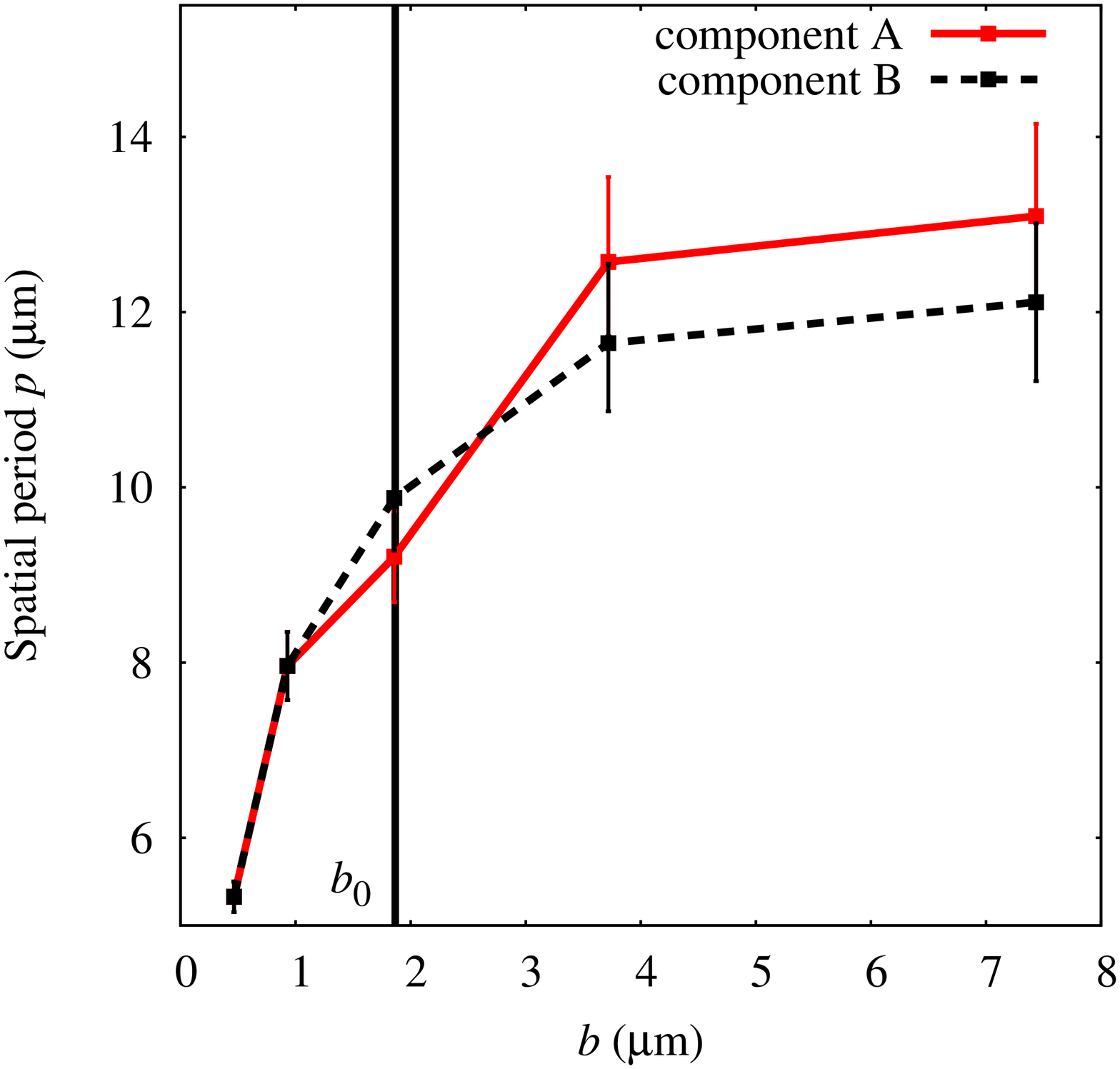}
\caption{Spatial period of Faraday waves as a function of inhomogeneity $b$ in the case of a symbiotic configuration for $\omega = \omega_\mathrm{F}$,
obtained using FFT analysis.}
\label{250-period}
\end{center}
\end{figure}

In order to estimate onset times for the emergence of resonant waves during real-time evolution, in figure \ref{160-ws-vis}
we plot time dependence of the pattern visibility function,
a dimensionless quantity defined according to \cite{F1} as the standard deviation of the radially integrated density profile from the corresponding ground state, normalized to the number of atoms. When visibility reaches the value of one, density patterns are clearly visible in the system. For weakly inhomogeneous collisions, the resonant waves emerge after around 80~ms, while for strongly inhomogeneous
collisions the resonant waves are smoother and visible clearly
only after 350-400~ms, which can be also seen from figure \ref{160-ws-psi12} upon closer inspection.
We observe a softening of
resonant waves for strongly inhomogeneous collisions, and their onset
time is considerably longer than that observed for weakly
inhomogeneous collisions. This can be attributed to a change in the shape of atomic clouds, since in this case the system becomes completely miscible,
as can be seen from figure \ref{Sym-Stationary}(e).

Figure \ref{160-period} shows the
spatial period of resonant waves for both components as a
function of the inhomogeneity parameter $b$. For strong inhomogeneity the period is almost the same for
both components, but as we decrease the strength of the
collisional inhomogeneity (i.e., increase $b$), they separate
out. This separation is a direct consequence of a different number
of atoms in components A and B. As in the case of a single-component condensate \cite{F8},
the spatial period of waves increases as the inhomogeneity weakens, and eventually saturates
to a value corresponding to the collisionally homogeneous case.

Figure \ref{250-ws-psi12} shows the dynamics of 
radially integrated longitudinal density profiles in the case of weak and strong collisional
inhomogeneity that illustrate the emergence of Faraday waves for a modulation frequency $\omega
= \omega_\mathrm{F}$. According to figure \ref{250-ws-vis}, the Faraday waves for weak collisional inhomogeneity emerge
after  150-200~ms, while for strong collisional
inhomogeneity they are clearly visible after 350-400~ms. Again we observe a long delay in the onset of Faraday waves for strong
collisional inhomogeneity, as in the resonant case, due to a change in the shape of atomic clouds.
Figure \ref{250-period} shows
the spatial period of Faraday waves as a function of the inhomogeneity parameter $b$ and we
again see that the periods separate out as we decrease the strength of the inhomogeneity (i.e.,
increase the value of $b$). The saturation appears again for weak inhomogeneity and the periods eventually
converge to their values in the case of homogeneous interactions.

\subsection{Segregated state}

\begin{figure}[!ht]
\begin{center}
\includegraphics[width=70 mm]{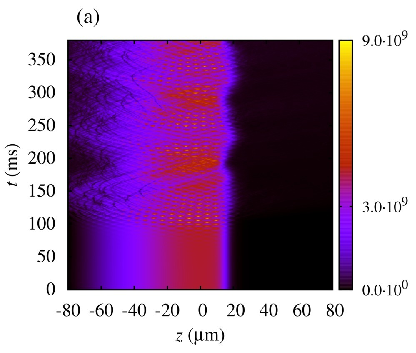}
\includegraphics[width=70 mm]{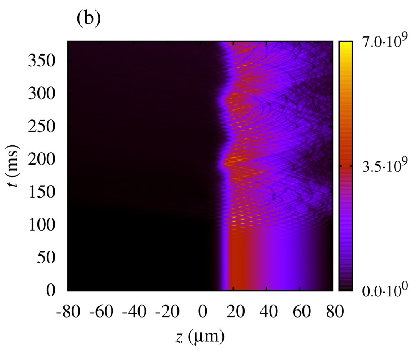}\\
\includegraphics[width=70 mm]{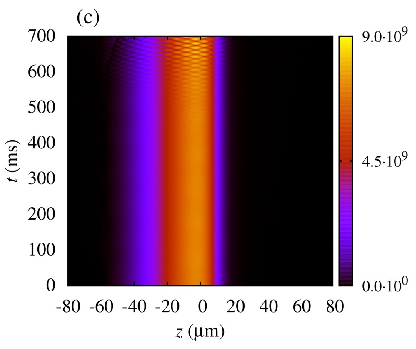}
\includegraphics[width=70 mm]{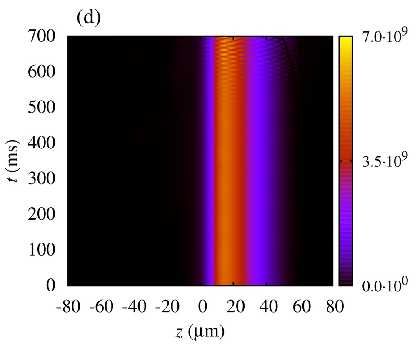}\\
\includegraphics[width=70 mm]{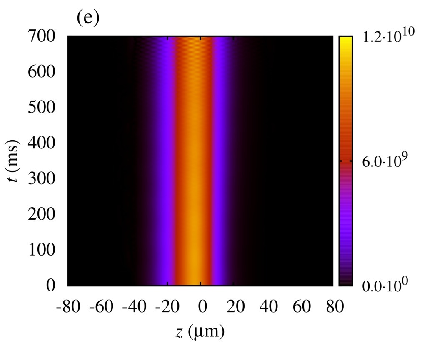}
\includegraphics[width=70 mm]{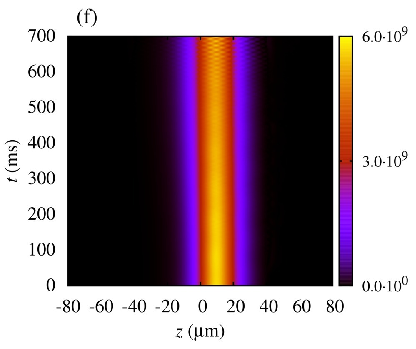}
\caption{Real-time evolution of radially integrated density profiles for weakly and strongly inhomogeneous interactions in the case of a segregated configuration
for $\omega = \omega_\mathrm{res}$. The panels on the left (right) correspond to the component A (B) for the inhomogeneity parameter
values: (a) and (b) $b = 4 b_0$, (c) and (d) $b = b_0/2$, (e) and (f) $b = b_0/4$. Note the softening of nonlinear excitations  as $b$ decreases and the system reaches an effectively linear regime.}
\label{160-segws-psi12}
\end{center}
\end{figure}

\begin{figure}[!ht]
\begin{center}
\includegraphics[width=70 mm]{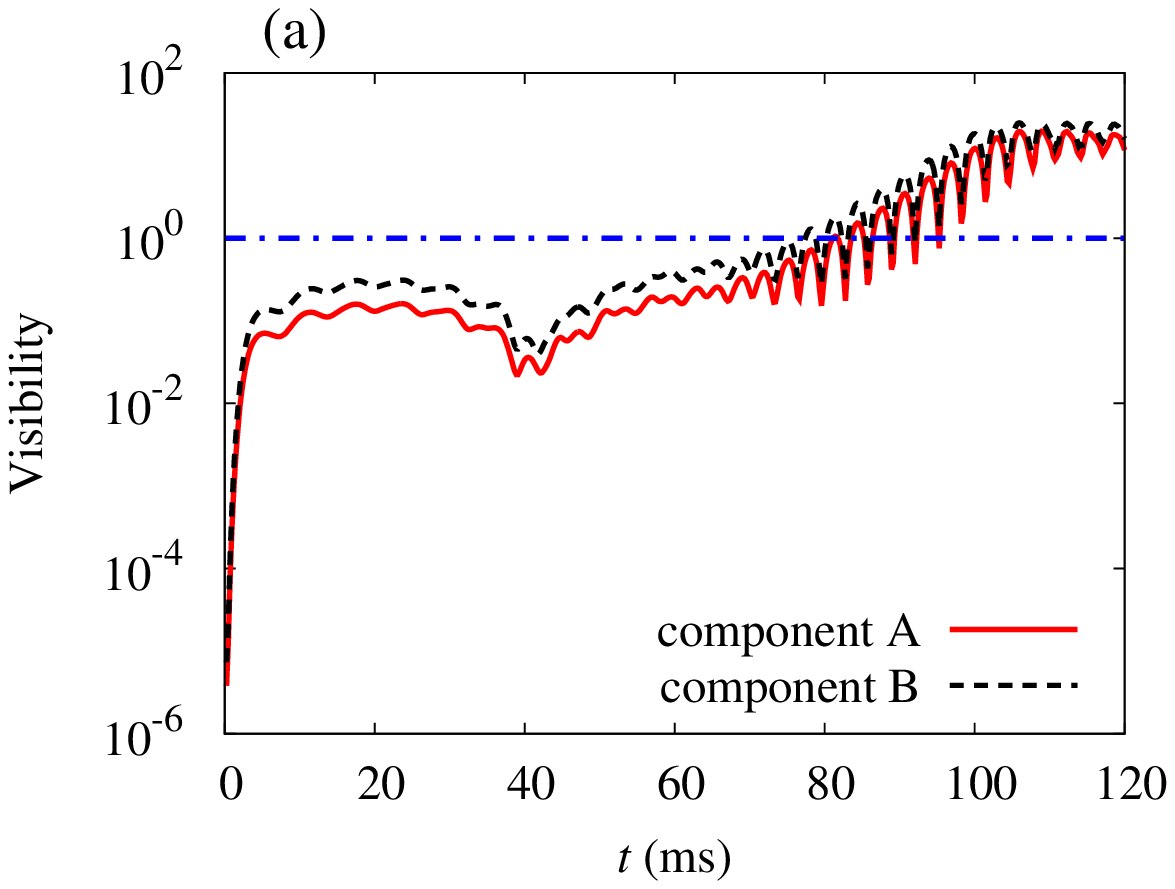}
\includegraphics[width=70 mm]{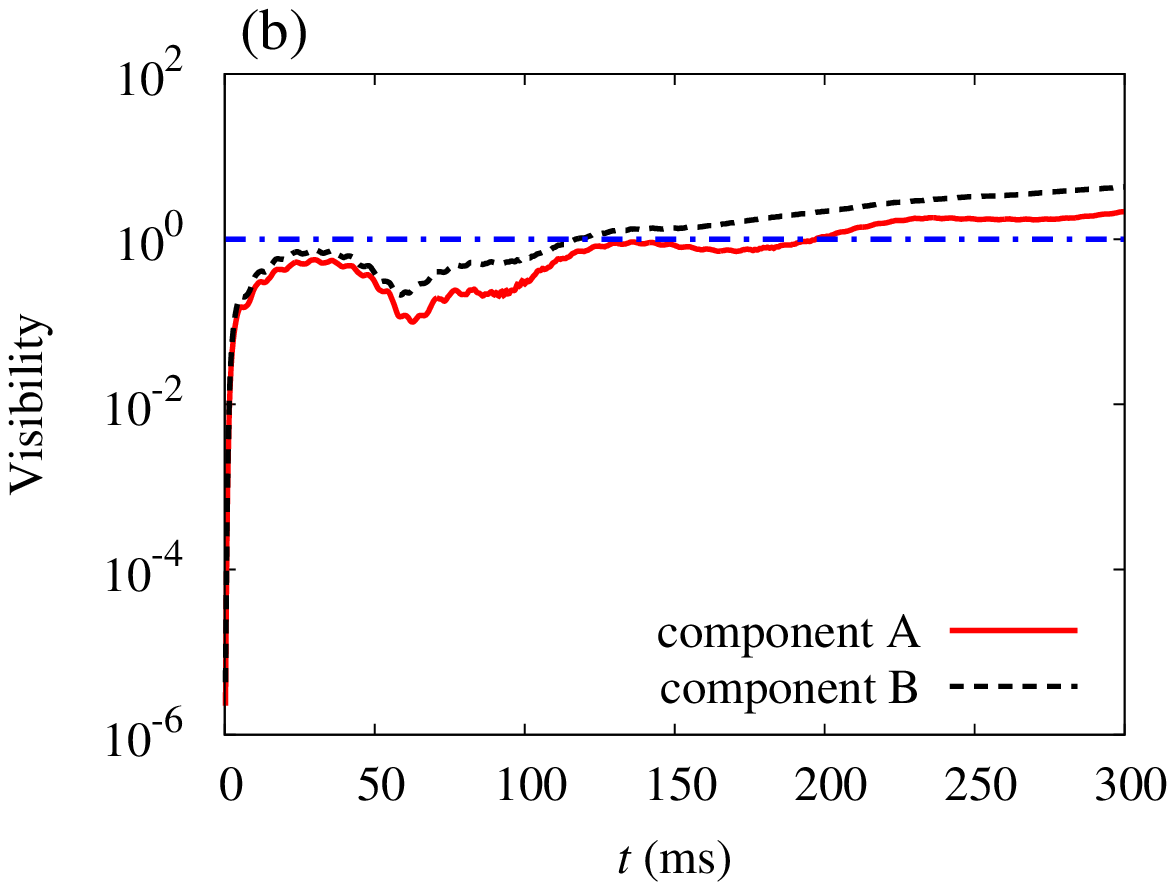}
\caption{Time dependence of the visibility during real-time evolution for weakly and strongly inhomogeneous interactions in the case of a segregated configuration
for $\omega = \omega_\mathrm{res}$ for the inhomogeneity parameter values: (a) $b = 4 b_0$, (b) $b = b_0/4$. The horizontal dashed-dotted line corresponds to the
visibility equal to one and denotes the onset of resonant waves.}
\label{160-segws-vis}
\end{center}
\end{figure}

\begin{figure}[!ht]
\begin{center}
\includegraphics[width=70 mm]{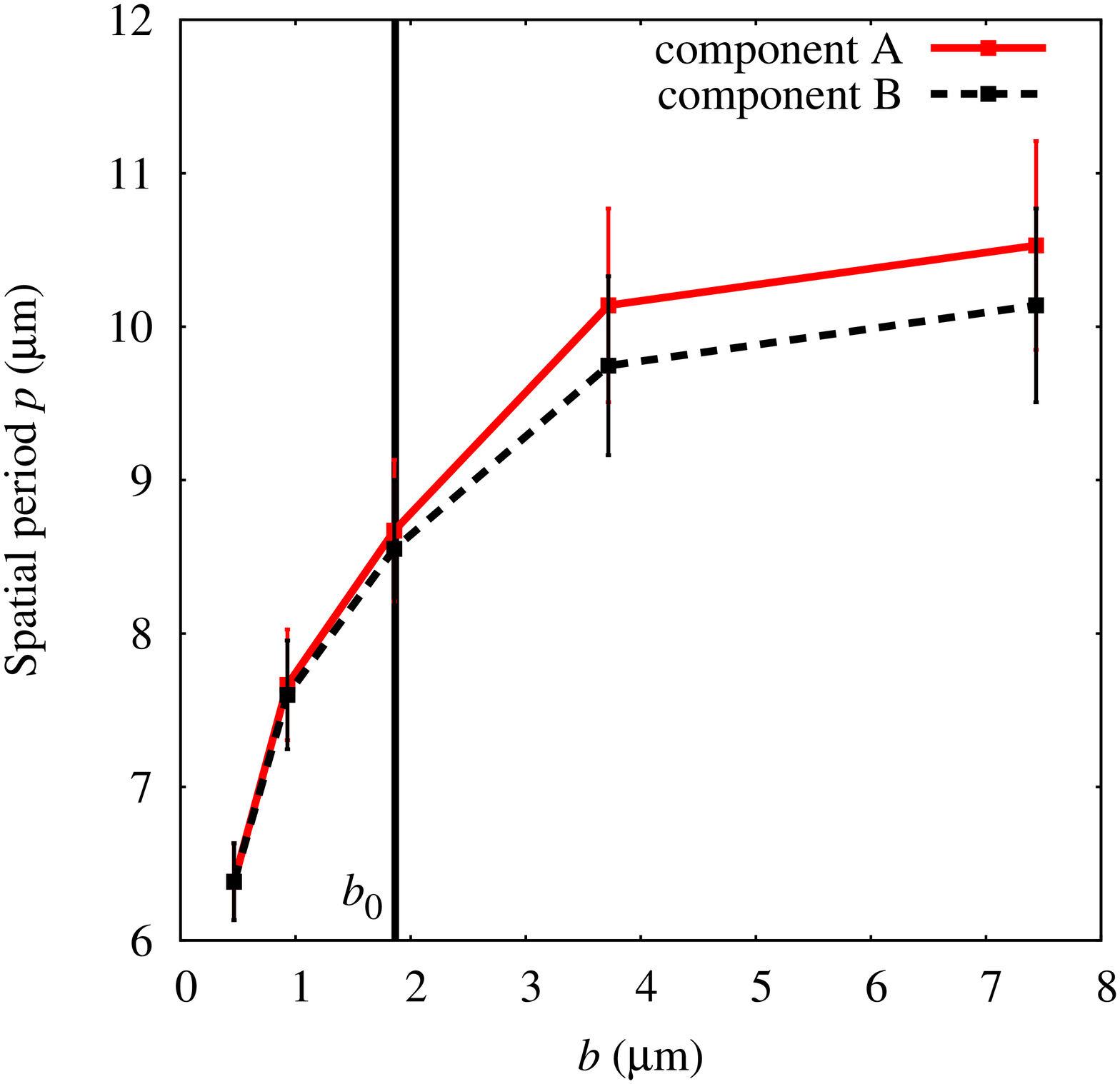}
\caption{Spatial period of resonant waves as a function of inhomogeneity $b$ in the case of a segregated configuration for $\omega = \omega_\mathrm{res}$,
obtained using FFT analysis.}
\label{160-seg}
\end{center}
\end{figure}

Here we discuss the dynamical evolution due to harmonic modulation of the radial part of the trapping potential for the case of a segregated initial system configuration.
Starting from the stationary solutions presented in section \ref{sec:seg-stationary}, figure \ref{160-segws-psi12} shows the resonant waves obtained for a modulation frequency $\omega =  \omega_\mathrm{res}$. In this case, for the segregated initial configuration, the resonant waves are clearly visible immediately after 80-100~ms for weak collisional inhomogeneity, while in the case of a strong collisional inhomogeneity, they require around 200~ms to develop and emerge, as can be seen in figure \ref{160-segws-vis}. Although the onset time increases with the inhomogeneity, it is not that pronounced as in the case of a symbiotic configuration. This can be attribited to the fact that the change in the atomic clouds shape in figure \ref{Seg-Stationary} is not that drastic as in figure \ref{Sym-Stationary}. Figure \ref{160-seg} shows the spatial periods of the resonant wave as a function of the inhomogeneity parameter $b$. We note the similar behaviour as in previous cases.

\begin{figure}[!ht]
\begin{center}
\includegraphics[width=70 mm]{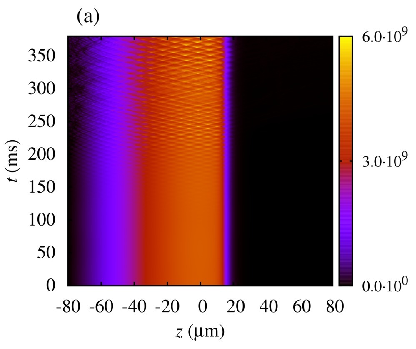}
\includegraphics[width=70 mm]{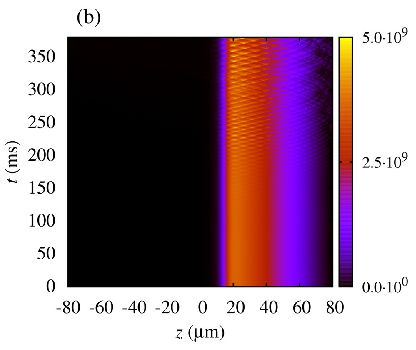}\\
\includegraphics[width=70 mm]{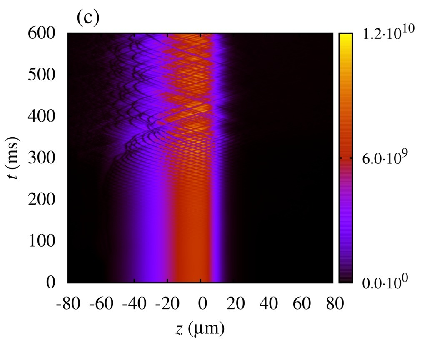}
\includegraphics[width=70 mm]{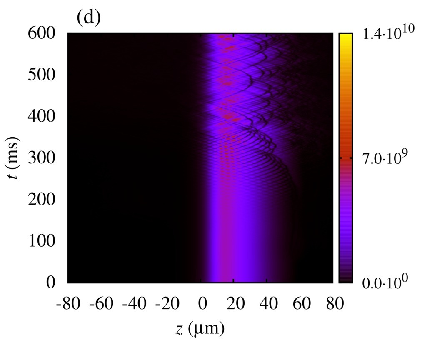}\\
\includegraphics[width=70 mm]{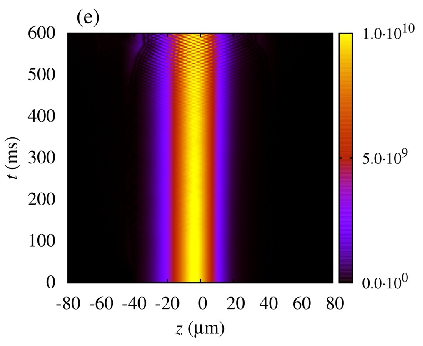}
\includegraphics[width=70 mm]{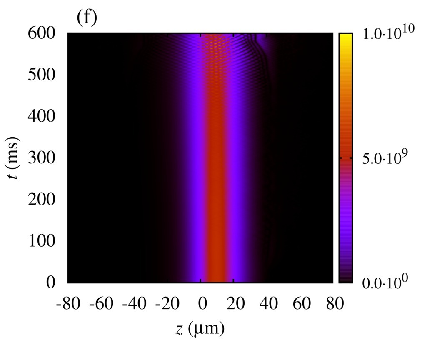}
\caption{Real-time evolution of radially integrated density profiles for weakly and strongly inhomogeneous interactions in the case of a segregated configuration
for $\omega = \omega_\mathrm{F}$. The panels on the left (right) correspond to the component A (B) for the inhomogeneity parameter
values: (a) and (b) $b = 4 b_0$, (c) and (d) $b = b_0/2$, (e) and (f) $b = b_0/4$. Note the softening of nonlinear excitations  as $b$ decreases and the system reaches an effectively linear regime}
\label{250-segws-psi12}
\end{center}
\end{figure}

\begin{figure}[!ht]
\begin{center}
\includegraphics[width=70 mm]{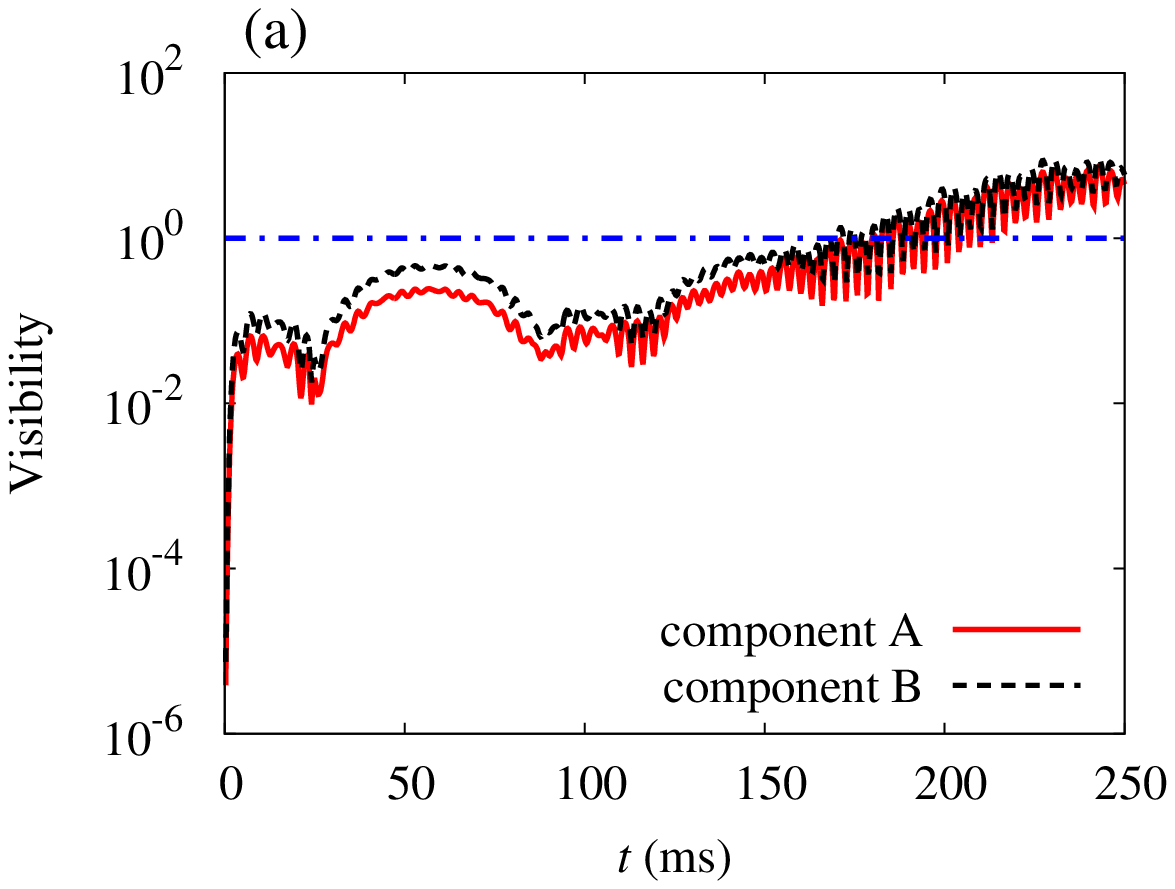}
\includegraphics[width=70 mm]{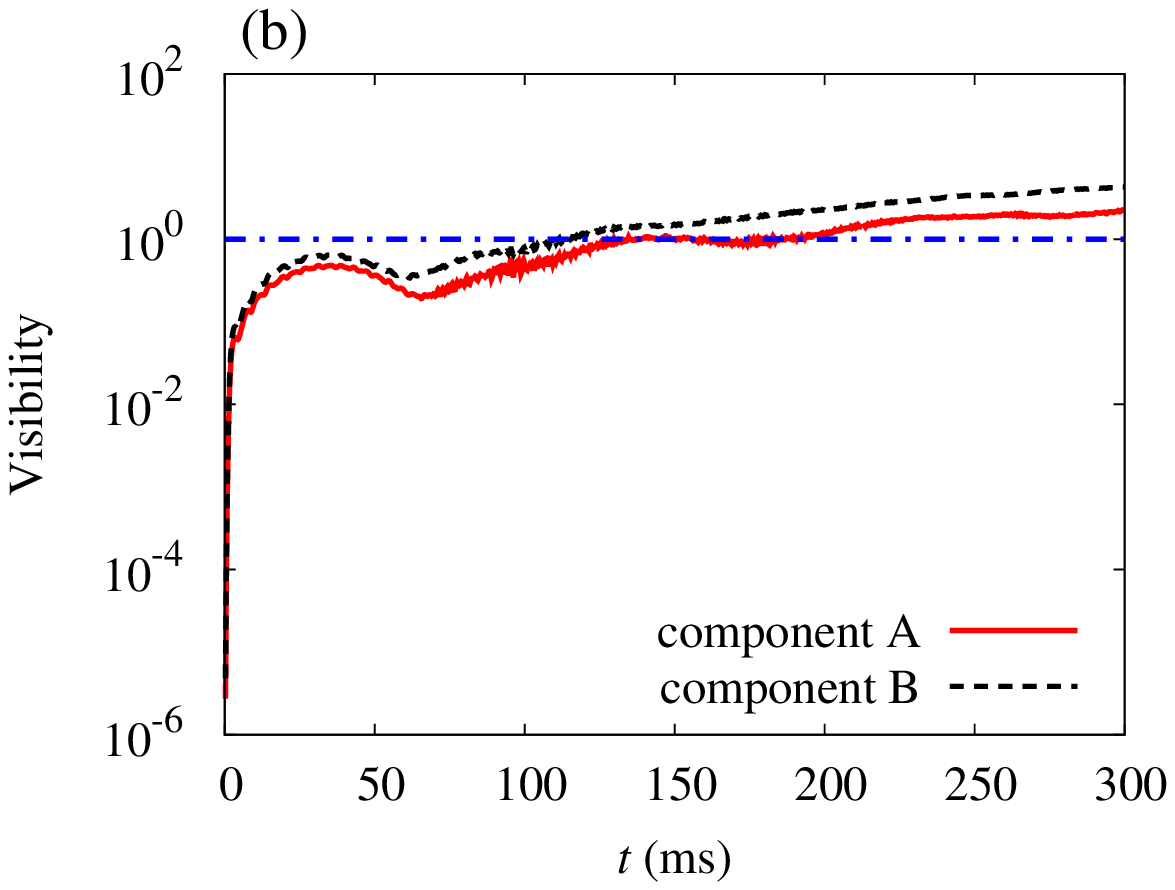}
\caption{Time dependence of the visibility during real-time evolution for weakly and strongly inhomogeneous interactions in the case of a segregated configuration
for $\omega = \omega_\mathrm{F}$ for the inhomogeneity parameter values: (a) $b = 4 b_0$, (b) $b = b_0/4$. The horizontal dashed-dotted line corresponds to the
visibility equal to one and denotes the onset of Faraday waves.}
\label{250-segws-vis}
\end{center}
\end{figure}

\begin{figure}[!ht]
\begin{center}
\includegraphics[width=70 mm]{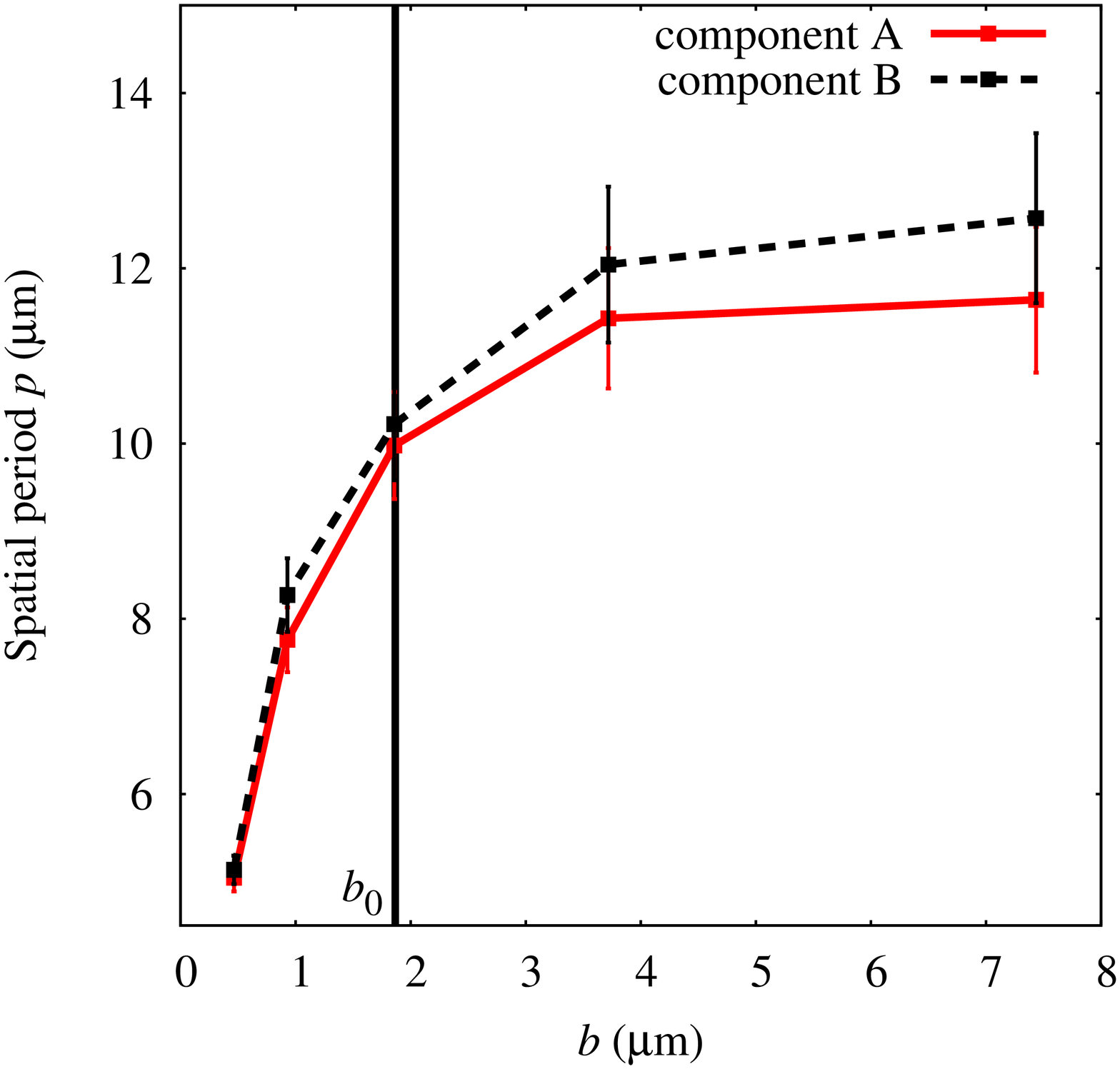}
\caption{Spatial period of Faraday waves as a function of inhomogeneity $b$ in the case of a segregated configuration for $\omega = \omega_\mathrm{F}$,
obtained using FFT analysis.}
\label{250-seg}
\end{center}
\end{figure}

The next set of graphs in figure \ref{250-segws-psi12} presents results for the non-resonant modulation frequency, $\omega=\omega_\mathrm{F}$.
As can be seen in figure \ref{250-segws-vis}, the emergence time of the Faraday waves again varies as the inhomogeneity is changed.
For weak collisional inhomogeneity, the Faraday waves are visible after around 150-200~ms, i.e., twice as much than for the resonant modulation. In the case of strong collisional inhomogeneity, the Faraday waves emerge after 200-250~ms, which represents only a slight increase, due to a same reason as in the resonant case.

Finally, figure \ref{250-seg} shows the spatial period of Faraday waves as a function of the inhomogeneity control parameter $b$.
If we compare figures \ref{160-period}, \ref{250-period} and figures \ref{160-seg}, \ref{250-seg}, we see that the spatial period of density waves in
both components are quite similar for weak spatial inhomogeneity for both symbiotic and segregated states, respectively.

\section{Conclusions}
\label{conclusions}

Summing up, we have shown through extensive numerical simulations
that binary condensates in the so-called collisionally
inhomogeneous regime  can reach an effectively linear regime in
which nonlinear effects fade out as the collisions become
localised at the centre of the magnetic trap. This behaviour is
independent of their intrinsic configuration (i.e., either
symbiotic or segregated) and can be most easily seen in the
increase of the instability onset times of Faraday and resonant
waves, and the transition to miscibility. Moreover, we have observed that, 
in addition to longer instability onset times, the spatial periods of resonant and
Faraday waves decrease as the inhomogeneity becomes stronger.
To excite resonant waves we used a driving frequency equal to the radial
frequency of the trap, while for Faraday waves we used an off-resonance
driving frequency which is in-between the radial frequency of trap and its first harmonic.

We stress that in reaching the aforementioned effectively linear regime,
the two topologically different configurations, the segregated and the symbiotic ones, exhibit some quantitative differences.
The changes in the spatial structure of the two-body interactions impact more significantly the symbiotic states than the segregated ones.
Inspecting figures \ref{160-ws-vis}, \ref{250-ws-vis}, \ref{160-segws-vis}, and \ref{250-segws-vis}, one notices that, close to the linear regime,
both the Faraday and the resonant waves emerge slower for symbiotic states than for segregated ones due to a longer instability
onset time and stronger fluctuations in the pattern visibility function.
This should be contrasted with the regime of homogeneous nonlinearity in which the two types of waves have very similar instability onset times.

We also observe that both stationary immiscible configurations typical for binary cigar-shaped
condensates with constant short-range interactions gradually turn
into a perfectly miscible configuration as the spatial profile of
the scattering length gets closer to a delta-function, with the
wave functions of the two components reaching an almost Gaussian-like
functions. This suggests that the inhomogeneity of the binary 
collisions can be used experimentally as an efficient tool to control 
the level of miscibility in the configurations of two-species BECs.

\ack
A.~I.~N. acknowledges support from ANCSI through project PN 16420202/2016,
and M.~C.~R. from CNCS-UEFISCDI under project PN-II-ID-PCE-2011-3-0972, while A.~B. was supported by the
Serbian Ministry of Education, Science, and Technological Development
under projects ON171017, OI1611005,  and IBEC, and by
the DAAD-German Academic and Exchange Service under project IBEC.
Numerical simulations were run in part on the PARADOX supercomputing facility at the Scientific Computing Laboratory of the Institute of Physics Belgrade.
J.~B.~S. acknowledges support by the Department of Science and Technology (DST) and Council of Scientific and Industrial
Research (CSIR) of the Government of India. R.~R.
acknowledges support by DST (Ref.~No: SR/S2/HEP-26/2012), CSIR
(Ref.~No: 03(1323)/14/EMR-II dated 03.~11.~2014), and
Department of Atomic Energy - National Board of Higher
Mathematics (Ref.~No: 2/48(21)/2014/NBHM(R.P.)/R \& D II/15451).

\section*{Appendix}
All presented numerical results were obtained by solving the corresponding coupled Gross-Pitaevskii equations in three spatial dimensions,
which effectively reduce to a two-dimensional problem due to cylindrical symmetry of the system.
The equations are solved using the split-step Crank-Nicolson semi-implicit method and cylindrically-symmetric version of
numerical programs available in \cite{GP1,GP2,GP3,GP4,GP5,GP6}. The size of the spatial grid was $2000\times 2000$ and grid steps were $h_\rho=0.002$ in the radial and $h_z=0.04$ in the longitudinal direction, expressed in terms of the longitudinal oscillator length $\ell=\sqrt{\hbar/(m\Omega_z)}=4.07~\mu$m. The time step used for real-time propagation was $\Delta t=5\times 10^{-5}/\omega_z=1.14~\mu$s. The stability of all numerical results was carefully checked and these discretization parameters were found to be sufficiently small to ensure full reliability and reproducibility of the obtained results.

Using the numerically obtained results for the radially integrated density profiles during the real-time evolution of the system with harmonically modulated radial trapping frequency, we have calculated spatial periods of resonant and Faraday density patterns approximately 50~ms after their emergence. Since driving of the system eventually destroys it due to continual pumping of the energy into it, the presented results for spatial periods correspond to behaviour of the system during experimentally relevant 100-200~ms after the onset of density patterns. Due to violent dynamics in the resonant case, numerically calculated spatial periods of density patterns in such a way may be valid in a shorter time-frame.


\section*{References}
\begin{thebibliography}{10}

\bibitem{PatternFormationReview}
Cross M C and Hohenberg P C 1993 \RMP {\bf 65} 851

\bibitem{St1}
Staliunas K, Longhi S and de Valc\'{a}rcel G J 2002 \PRL {\bf 89} 210406

\bibitem{St2}
Staliunas K, Longhi S and de Valc\'{a}rcel G J 2004 \PR A {\bf 70} 011601(R)

\bibitem{Da1}
Kramer M, Tozzo C and DalfovoF 2005 \PR A {\bf 71} 061602(R)

\bibitem{Da2}
Modugno M, Tozzo C and Dalfovo F 2006 \PR A {\bf 74} 061601(R)\\
Nicolin A I, Jensen M H and Carretero-Gonzalez R 2007 \PR E {\bf 75} 036208\\
Achilleos V, Frantzeskakis D J, Kevrekidis P G, Schmelcher P and Stockhofe J 2015 {\it Rom. Rep. Phys.}  {\bf 67} 235\\
Vardi A 2015 {\it Rom. Rep. Phys.} {\bf 67} 67

\bibitem{ReviewRRP}
Bagnato V S, Frantzeskakis D J, Kevrekidis P G, Malomed B A and Mihalache D 2015 {\it Rom. Rep. Phys.} {\bf 67} 5

\bibitem{RRRRP}
Radha R and  Vinayagam P S 2015 {\it Rom. Rep. Phys} {\bf 67} 147

\bibitem{F1}
Engels P, Atherton C and Hoefer M A 2007 \PRL {\bf 98} 095301

\bibitem{F2}
Abe H, Ueda T, Morikawa M, Saitoh Y, Nomura R and Okuda Y 2007 \PR E {\bf 76} 046305

\bibitem{F3}
Ueda T, Abe H, Saitoh Y, Nomura R and Okuda Y 2007 {\it J. Low Temp. Phys} {\bf 148} 553

\bibitem{F4}
Pollack S E, Dries D, Hulet R G, Magalhaes K M F, Henn E A L, Ramos E R F, Caracanhas M A and Bagnato V S 2010 \PR A {\bf 81} 053627

\bibitem{F5}
Nicolin A I, Carretero-Gonzalez R and Kevrekidis P G 2007 \PR A {\bf 76} 062609

\bibitem{F6}
Nicolin A I 2011 {\it Rom. Rep. Phys.} {\bf 63} 1329

\bibitem{F7}
Nath R and Santos L 2010 \PR A {\bf 81} 033626

\bibitem{F8}
Bala\v z A, Paun R, Nicolin A I, Balasubramanian S and Ramaswamy R 2014 \PR A {\bf 89} 023609

\bibitem{F9}
Nicolin A I, Bala\v z A, Sudharsan J B and Radha R 2014 {\it Rom. J. Phys.} {\bf 59} 204

\bibitem{F11}
Tang R A, Li H C and Xue J K 2011 \jpb {\bf 44} 115303

\bibitem{F12}
Staliunas K 2011 \PR A {\bf 84} 013626

\bibitem{Dis1}
Nikoli\' c B, Bala\v z A and Pelster A 2013 \PR A {\bf 88} 013624

\bibitem{Dis2}
Khellil T and Pelster A 2016 {\it J. Stat. Mech.-Theory Exp.} 063301

\bibitem{Dis3}
Khellil T, Bala\v z A and Pelster A 2016 \NJP {\bf 18} 063003

\bibitem{QF1}
Lima A R P and Pelster A 2011 \PR A {\bf 84} 041604(R)

\bibitem{QF2}
Lima A R P and Pelster A 2012 \PR A {\bf 86} 063609\\
Bogojevi\'c A, Bala\v z A and Beli\'c A 2005 \PR E {\bf 72} 036128\\
Bogojevi\'c A, Vidanovi\'c I, Bala\v z A and Beli\'c A 2008 {\it Phys. Lett.} A {\bf 372} 3341\\
Vidanovi\'c I, Bogojevi\'c A, Bala\v z A and Beli\'c A 2009 \PR E {\bf 80} 066706\\
Bala\v z A, Bogojevi\'c A, Vidanovi\'c A and Pelster A 2009 \PR E {\bf 79} 036701\\
Bala\v z A, Vidanovi\'c I, Bogojevi\'c A and Pelster A 2010 {\it Phys. Lett.} A {\bf 374} 1539\\
Al-Jibbouri H, Vidanovi\'c I, Bala\v z A and Pelster A 2013 \jpb {\bf 46} 065303

\bibitem{CIC}
Theocharis G, Schmelcher P, Kevrekidis P G and Frantzeskakis D J 2005 \PR A {\bf 72} 033614

\bibitem{OpticalFeshbach}
Yamazaki R, Taie S, Sugawa S and Takahashi Y 2010 \PRL {\bf 105} 050405

\bibitem{OpticalFeshbach2}
Dong G, Hu B and Lu W 2006 \PR A {\bf 74} 063601

\bibitem{verhaar}
Verhaar B J, van Kempen E G M and Kokkelmans S J J M F 2009 \PR A {\bf 79} 032711

\bibitem{middlekamp}
Middelkamp S, Chang J J, Hamner C, Carretero-Gonzalez R, Kevrekedis P G, Achilleos V, Frantzeskakis D J, Schmelcher P and Engels P 2011 {\it Phys. Lett.} A {\bf 375} 642

\bibitem{hamner}
Hamner C, Chang J J, Engels P and Hoefer M A 2011 \PRL {\bf 106} 065302

\bibitem{ImaginaryTimeAlgorithms}
Chiofala M L, Succi S and Tosi M P 2000 \PR E {\bf 62} 7438

\bibitem{ExplicitFiniteDifferenceScheme}
Cerimele M M, Chiofalo M L, Pistella F, Succi S and Tosi M P 2000 \PR E {\bf 62} 1382

\bibitem{TimeSplittingSpectralMethods}
Bao W, Jaksch D and Markowich P A 2003 {\it J. Comput. Phys.} {\bf  187} 318

\bibitem{BasisSet}
Tiwari R P and Shukla A 2006 {\it Comput. Phys. Commun.} {\bf 174} 966

\bibitem{Symplectic}
Hua W, Liu X and Ding P 2006 {\it J. Math. Chem.} {\bf 40} 243

\bibitem{GP1}
Muruganandam P and Adhikari S K 2009 {\it Comput. Phys. Commun.} {\bf 180} 1888

\bibitem{GP2}
Vudragovi\' c D, Vidanovi\' c I, Bala\v z A, Muruganandam P and Adhikari S K 2012 {\it Comput. Phys. Commun.} {\bf 183} 2021

\bibitem{GP3}
Kishor Kumar R, Young-S. L-E, Vudragovi\' c D, Bala\v z A, Muruganandam P and Adhikari S K 2015 {\it Comput. Phys. Commun.} {\bf 195} 117

\bibitem{GP4}
Lon\v car V, Bala\v z A, Bogojevi\' c A, \v Skrbi\' c A, Muruganandam P and Adhikari S K 2016 {\it Comput. Phys. Commun.} {\bf 200} 406

\bibitem{GP5}
Satari\' c B, Slavni\' c V, Beli\' c A, Bala\v z A, Muruganandam P and Adhikari S K 2016 {\it Comput. Phys. Commun.} {\bf 200} 411

\bibitem{GP6}
Young-S. L-E, Vudragovi\' c D, Muruganandam P, Adhikari S K and Bala\v z A 2016 {\it Comput. Phys. Commun.} {\bf 204} 209

\bibitem{miscibility}
Tin-Lun H and Shenoy V B 1996 \PRL {\bf 77} 3276

\bibitem{ref50A}
Ao P and Chui S T  1998 \PR A {\bf 58} 4836

\bibitem{ref38A}
Verhaar B J, van Kempen E G M and Kokkelmans S J J M F 2009 \PR A {\bf 79} 032711

\bibitem{Ivana}
Vidanovi\' c I, van Druten N J and Haque M 2013 \NJP {\bf 15} 035008

\bibitem{FW2012}
Bala\v z A and Nicolin A I 2012 \PR A {\bf 85} 023613


\end{thebibliography}
\end{document}